\documentclass[prd,notitlepage,showpacs,amsmath,amssymb,preprintnumbers,nofootinbib,longbibliography]{revtex4-1}
\usepackage{graphicx}
\usepackage{multirow}
\usepackage{amsmath}
\usepackage{amssymb}
\usepackage{xspace}
\usepackage{gensymb}
\usepackage{hhline}
\usepackage{multirow}
\usepackage{booktabs}
\usepackage{hyperref}
\usepackage{cleveref}
\usepackage{aas_macros}
\usepackage{soul}
\usepackage{xcolor}
\usepackage{array}
\usepackage{adjustbox}
\usepackage{cancel}
\usepackage{appendix}

\newlength{\wdo}
\newcommand{\stroke}[1]{{$#1$}%
\settowidth{\wdo}{${#1}$} {\kern-\wdo}%
\partialvartstrokedint}

\makeatletter
\newcommand{\fancysep}{%
  \@afterindentfalse
  {\begin{center}
    \resizebox{0.8\linewidth}{0.4ex}{{%
        \fontsize{20}{24}\usefont{U}{webo}{xl}{n}{4}}}%
  \end{center}}\@afterheading}
\makeatother

{\vskip 1.0em
\framebox[\columnwidth][r]{%
\begin{minipage}[c]{\columnwidth}%
\vspace{-1.0em}%
#1%
\end{minipage}}}
{\vskip 1.0em}

\def\XXint#1#2#3{{\setbox0=\hbox{$#1{#2#3}{\int}$}
     \vcenter{\hbox{$#2#3$}}\kern-.5\wd0}}

\newcommand{\beq}{\begin{equation}}
\newcommand{\eeq}{\end{equation}}
\newcommand{\beqa}{\begin{eqnarray}}
\newcommand{\eeqa}{\end{eqnarray}}

\newcommand{\eps}{\ensuremath{\epsilon}\xspace}

\newcommand{\mpl}{\ensuremath{m_\textnormal{pl}}\xspace}

\newcommand{\neff}{\ensuremath{N_\textnormal{eff}}\xspace}
\newcommand{\lcdm}{\ensuremath{\Lambda}CDM\xspace}

\newcommand{\summnu}{\ensuremath{\Sigma m_\nu}\xspace}

\newcommand{\gstar}{\ensuremath{g_\star}\xspace}

\newcommand{\ben}{\begin{enumerate}}
\newcommand{\een}{\end{enumerate}}

\newcommand{\yp}{\ensuremath{Y_{\rm P}}\xspace}
\newcommand{\dtoh}{\ensuremath{{\rm D/H}}\xspace}

\newcommand{\rhorad}{\ensuremath{\rho_{\rm rad}}\xspace}

\newcommand{\gstardec}{\ensuremath{g_{\star}^{\rm dec}}\xspace}
\newcommand{\gstarsdec}{\ensuremath{g_{\star,S}^{\rm dec}}\xspace}
\newcommand{\kfs}{\ensuremath{k_{\rm fs}}\xspace}
\newcommand{\kfsz}{\ensuremath{k_{\rm fs,0}}\xspace}
\newcommand{\vth}{\ensuremath{v_{\rm th}}\xspace}

\begin{document}

\title{Implications on Cosmology from Dirac Neutrino Magnetic Moments}

\author{E. Grohs$^{1,2}$}
\author{A. B. Balantekin$^3$}

\affiliation{$^{1}$Department of Physics, University of California Berkeley, Berkeley, California
94720, USA}
\affiliation{$^{2}$Department of Physics, North Carolina State University,
Raleigh, North Carolina 27607, USA}
\affiliation{$^{3}$Department of Physics, University of
Wisconsin-Madison, Madison, Wisconsin 53706, USA}

\date{\today}

\begin{abstract}

The mechanism for generating neutrino masses remains a puzzle in particle
physics.  If neutrino masses follow from a Dirac mass term, then neutrino
states exist with opposite chirality compared to their weakly-interacting
counterparts.  These inactive states do not interact with their active
counterparts at measurable scales in the standard model.  However, the
existence of these states can have implications for cosmology as they
contribute to the radiation energy density at early times, and the matter energy density at late times.  How Dirac neutrinos may populate thermal states via an anomalous magnetic moment
operator is the focus of this work.  A class of models where all neutrinos have a magnetic
moment independent of flavor or chirality is considered.  Subsequently, the cross sections for neutrinos
scattering on background plasma particles are calculated so that the relic inactive
neutrino energy is derived as a function of plasma temperature.  To do so, one needs cross sections for scattering on all electrically charged
standard-model particles.  Therefore, the scattering cross section
between a neutrino and $W$-boson via the magnetic moment vertex is derived.  Current
measurements put a constraint on the size of the neutrino magnetic moment from
the cosmological parameter \neff and light-element primordial abundances.
Finally, how the extra Dirac states contribute to the matter
energy density at late times is investigated by examining neutrino free-streaming.

\end{abstract}

\maketitle

\section{Introduction}

In his ``Dear Radioactive Ladies and Gentlemen'' letter to the T\"ubingen meeting of the German Physical Society (reproduced in Ref.\ \cite{Pauli:1930pc}) Wolfgang Pauli, in addition to  proposing the existence of neutrino itself, implied that the neutrino is massive and hence it interacts via its magnetic dipole moment. Since Pauli did not explore the possibility of a new kind of interaction, i.e., the weak interaction, the magnetic moment he could deduce was too large. After the weak interaction was introduced by Enrico Fermi and it was realized that neutrinos could be massless, interest in the electromagnetic interactions of neutrinos waned since symmetry considerations suggest that neutrino magnetic moment would vanish for massless neutrinos. Earlier surveys did not come up with any  experimental evidence for electromagnetic interactions of neutrinos \cite{Bernstein:1963qh}. 
However, as solar neutrino experiments found increasingly strong evidence for the presence of non-zero neutrino masses, papers exploring astrophysical and cosmological implications of a non-zero neutrino magnetic moment started to appear in the literature \cite{Morgan:1981psa,Morgan:1981zy,Okun:1986hi,Okun:1986na,Raffelt:1989xu,Lim:1987tk,Akhmedov:1988uk,Balantekin:1990jg,2002PhR...370..333D}. 
Indeed one of the solutions of the solar neutrino problem was to invoke interactions of neutrinos with solar magnetic fields \cite{Cisneros:1970nq}.

Within the standard model, neutrinos are taken to be massless.
If indeed neutrinos are massive, as the solar neutrino experiments suggested, the question arised as to how they obtain their masses in an extension to the standard model.
Since they are neutral fermions, a neutrino mass term added to the standard-model Lagrangian can produce a discernible difference between Dirac or Majorana character. A Dirac neutrino is distinct from its antiparticle: Dirac neutrinos carry lepton number $+1$ and Dirac antineutrinos carry lepton number $-1$. Conversely, a Majorana neutrino is identical to its antiparticle and consequently there is no conserved lepton number with Majorana neutrinos. A free Dirac neutrino, like all the other charged fermions, is described by a spinor with four independent components. In contrast, since a Majorana neutrino is its own antiparticle (i.e., equal to its charge-conjugate up to a phase), its spinor has only two independent components. A direct consequence of this is that a Majorana neutrino cannot have a diagonal (i.e., connecting two mass eigenstates which are the same) magnetic moment; but magnetic moments connecting two different mass eigenstates are permitted. Dirac neutrinos have no such constraint.

Determining the Dirac versus Majorana character of neutrinos is a major area of research and there exist many terrestrial experiments dedicated to this search, e.g., neutrinoless double beta decay \cite{NEXT:2015wlq,EXO:2017poz,CUORE:2019yfd,GERDA:2020xhi,LEGEND:2021bnm,SNO:2021xpa,KamLAND-Zen:2022tow,Majorana:2022udl}.
Complementary to the terrestrial searches, there has been a long history of using cosmology to probe neutrino properties and interactions. Early work by Schramm and his collaborators \cite{Steigman:1977kc,Gunn:1978gr} connecting the number of neutrinos to cosmological parameters and observables brought those efforts to the forefront. This work particularly emphasized using the effective relativistic degrees of freedom, $N_{\rm eff}$, and the neutrino mass density in the universe to constrain the neutrino parameters.
The interplay of terrestrial experiments and cosmological observations continues to the present day \cite{2022arXiv220307377G} and this work follows in the same spirit.

Only left-handed neutrinos (and right-handed antineutrinos)
\footnote{We note that these states should properly be called left-chiral or right-chiral, not left-handed or right-handed.  Nevertheless, we adopt the nomenclature present in the literature when referring to the chiral states.},
which are referred to as ``active'',  take place in weak interactions. Any neutral fermion which does not participate in weak interactions is ``sterile'', although this term is more frequently used for those neutral fermions which mix with the active states and have different mass eigenvalues.  
To avoid confusion in this work, we will label the opposite chirality Dirac states (right-handed neutrinos and left-handed antineutrinos) as ``inactive.''
Additional interactions of neutrinos beyond the weak interactions (such as electromagnetic couplings) allow neutrinos to remain in thermal contact longer during the Big Bang Nucleosythesis (BBN) epoch \cite{Morgan:1981psa}. Imposing the condition that production of inactive neutrino states does not alter the primordial $^4$He abundance Morgan obtained a limit of $\sim 10^{-11} \mu_B$ \cite{Morgan:1981zy} on the neutrino magnetic moment, where $\mu_B=e/(2m_e)$ is the Bohr magneton. Further imposing the condition that inactive states do not increase the effective relativistic degrees of freedom in excess of one more neutrino species, Morgan's limit was relaxed by a factor of $\sim 3$ \cite{Fukugita:1987uy}. 
This limit only applies to Dirac neutrinos since for Majorana neutrinos right-handed states are not additional neutrino states, but represent antineutrinos.  
The energy dependence of the reaction cross sections due to the contribution of the electromagnetic couplings of neutrinos is different than that of the usual weak interaction couplings. For Majorana neutrinos with transition (i.e., connecting two different mass eigenstates) magnetic moments such reactions convert neutrinos into antineutrinos and vice versa. The resulting change in the reaction rates would 
alter the way neutrinos decouple from the plasma of electrons/positrons and photons. Such considerations can be used to limit magnetic moments of Majorana neutrinos as was done in Ref.\  \cite{Vassh_tdecoup}. The purpose of this paper is to improve limits on the magnetic moments of Dirac neutrinos using a careful assessment of the physics of decoupling in the Early Universe, i.e., the epoch at which the scattering rates of inactive neutrinos become too small to maintain thermal equilibrium with the plasma of standard-model constituents. 

To determine the decoupling of the inactive neutrinos, we require a form for the electromagnetic interaction.
We introduce the electromagnetic vertex function to characterize electromagnetic interactions below electroweak symmetry breaking \cite{2004JETP...99..254D}
\begin{equation}\label{eq:form_factor}
  F_\alpha(k) = f_Q(k^2)\gamma_\alpha + f_M(k^2)i\sigma_{\alpha\beta}k^{\beta} -f_E(k^2)\sigma_{\alpha\beta}k^{\beta}\gamma_5 + f_A(k^2)(k^2\gamma_\alpha - k_\alpha\cancel{k})\gamma_5.
\end{equation}
In Eq.\ \eqref{eq:form_factor}, we adopt the conventions $\sigma_{\alpha\beta}=i[\gamma_\alpha,\gamma_\beta]/2$, $\gamma_5=-i\gamma^0\gamma^1\gamma^2\gamma^3$, and $\cancel{k}=\gamma_\alpha k^{\alpha}$.  $f_Q$, $f_M$, $f_E$, and $f_A$ are the electric monopole, magnetic dipole, electric dipole, and anapole form factors respectively, for momentum transfer $k_\alpha$.
Using the operator for the magnetic dipole interaction in Eq.\ \eqref{eq:form_factor}, we will calculate scattering amplitudes, cross sections, and rates as a function of plasma temperature $T$.
We calculate both elastic scattering ($\nu+c\leftrightarrow c+\nu$) and annihilation ($\nu+\overline{\nu}\leftrightarrow c + \overline{c}$) processes between neutrinos and charged particles $c$.  As a result of considering the scattering interactions at times {\em after} the ElectroWeak Transition (EWT), we take the Higgs and electroweak bosons as massive particles.
A corollary of this treatment is the inclusion of electromagnetic interactions between $W^\pm$ bosons and neutrinos, which we will show have profound effects on setting limits on the neutrino magnetic moment.
We will loosen the restriction for the Quark-Hadron Transition (QHT) -- where quark and gluon degrees of freedom disappear and are replaced by hadrons -- and consider epochs before and after this transition.
The QHT is included using the approximate treatment described in Appendix C and based off of Ref.\ \cite{2014PhRvC..90b4915A}. 

The outline of this paper is as follows.
We summarize neutrino magnetic moment interactions and the pertinent cosmology in Secs.\ \ref{sec:MM} and \ref{sec:cosmo}, respectively. Our results for the Early and Later Universe are given in Secs.\ \ref{sec:early} and \ref{sec:later}. In Section \ref{sec:conclusions} we present our conclusions. Appendices \ref{app:a} and \ref{app:b} cover description of differential cross sections with magnetic moment vertices and thermal averaging of the cross sections.  Appendix \ref{app:qht} details our treatment of the QHT. 
Throughout this work, we use natural units where $\hbar=c=k_B=1$.

\section{Magnetic Moments}
\label{sec:MM}

Comprehensive reviews of neutrino electromagnetic interactions in the context of both the Standard Model and physics beyond the Standard Model are available in the literature \cite{Broggini:2012df,Giunti:2014ixa,Balantekin:2018azf}.  
The value of the neutrino magnetic moment in the minimally-extended (i.e., to include the neutrino mass) standard electroweak theory is very small. Using the expression for the one-loop electromagnetic vertex for fermions \cite{Lee:1977tib} it was calculated to be order of $10^{-20} \mu_B$ \cite{Fujikawa:1980yx}. Given our current knowledge of the neutrino masses and mixing angles, the updated prediction of the Standard Model, minimally extended to allow massive neutrinos, for the electron neutrino magnetic moment is even smaller 
\cite{Balantekin:2013sda}. In contrast, the most stringent laboratory limit on the neutrino magnetic moment obtained from electron scattering experiments is orders of magnitude larger: $2.9 \times 10^{-11} \mu_B$ \cite{Beda:2012zz}. 
Recently excess electron recoil events at the XENON1T detector \cite{XENON:2020rca} was interpreted as a possible signature of the neutrino magnetic moment \cite{Miranda:2020kwy}. 
PandaX collaboration reports a neutrino magnetic moment limit of $4.9 \times 10^{-11} \mu_B$ using the low energy electron recoil events \cite{PandaX-II:2020udv}. A recent analysis of the LUX-ZEPLIN data similarly limits the effective neutrino magnetic moment data to be less than $1.1\times10^{-11} \mu_B$
\cite{AtzoriCorona:2022jeb}.
Finally, a recent analysis of XENONnT data yields the most stringent limit for electron-flavor neutrino magnetic moment of $0.9\times10^{-11} \mu_B$ \cite{2023PhLB..83937742K}.
All three limits would rule out the neutrino magnetic moment interpretation of the XENON1T data. 
 
Large magnetic moments of neutrinos would have very interesting implications for astrophysics and cosmology. If there is an electromagnetic channel to produce neutrinos besides the usual weak one, then these additional neutrinos transfer more of the energy and entropy over large distances. It was remarked quite some time ago that extra energy loss due to the additional electromagnetic neutrino pair emission can limit the value of neutrino magnetic moment \cite{Bernstein:1963qh}. Indeed right after the observation of SN 1987A, it was shown that  bounds on the flux of right-handed neutrinos from a core-collapse supernova can be translated into bounds on neutrino magnetic moments \cite{Lattimer:1988mf,Barbieri:1988nh,Notzold:1988kz}. Perhaps the tightest astrophysical bound comes from red giant stars at globular clusters;  
the increased energy loss resulting from the electromagnetic neutrino pair production near the helium flash could lead to an increased core mass 
\cite{Raffelt:1989xu}. The most recent such analysis yields a limit in the range of $(1.2-1.5) \times 10^{-12} \mu_B$ \cite{Capozzi:2020cbu}. Other energy loss arguments typically yield less stringent limits. For example additional energy losses would eliminate the blue loops in the evolution of intermediate-mass stars; hence for Cepheid stars to exist, the neutrino magnetic moment should be smaller than the range $\sim 2 \times 10^{-10} \mu_B - 4 \times 10^{-11} \mu_B$ \cite{Mori:2020niw}. Similarly if the neutrino magnetic moment is of the order of $10^{-12} \mu_B$, additional energy losses can explain the enhanced lithium abundance observed in red clump stars \cite{Mori:2020qqd}. An examination of the pulsations \cite{Corsico:2014mpa} or the luminosity function of hot white dwarfs \cite{2014A&A...562A.123M} give similar limits. 
However such limits are subject to large uncertainties, such as the rate of the $^{12}$C reaction or the stellar metallicity. 
It was suggested that it is possible to evade such astrophysical limits \cite{Babu:2020ivd} 
by invoking new interactions of the neutrino with a light scalar boson \cite{Ge:2018uhz,Smirnov:2019cae,Babu:2019iml}. 
One can use spin-flavor precession of neutrinos to assess the value of the neutrino magnetic moment.  A more recent analysis using this approach with ultra-high-energy neutrinos is consistent with a limit of $1.2\times10^{-11}\mu_B$ \cite{ALOK2023137791}.
More recent work from astrophysics and cosmology considered transition magnetic moments between active and additional sterile neutrino states \cite{Brdar:2020quo,Miranda:2021kre,Brdar:2023tmi}. These limits are also of the order of $10^{-11} \mu_B$. 

\begin{figure}
  \includegraphics[width=0.5\columnwidth]{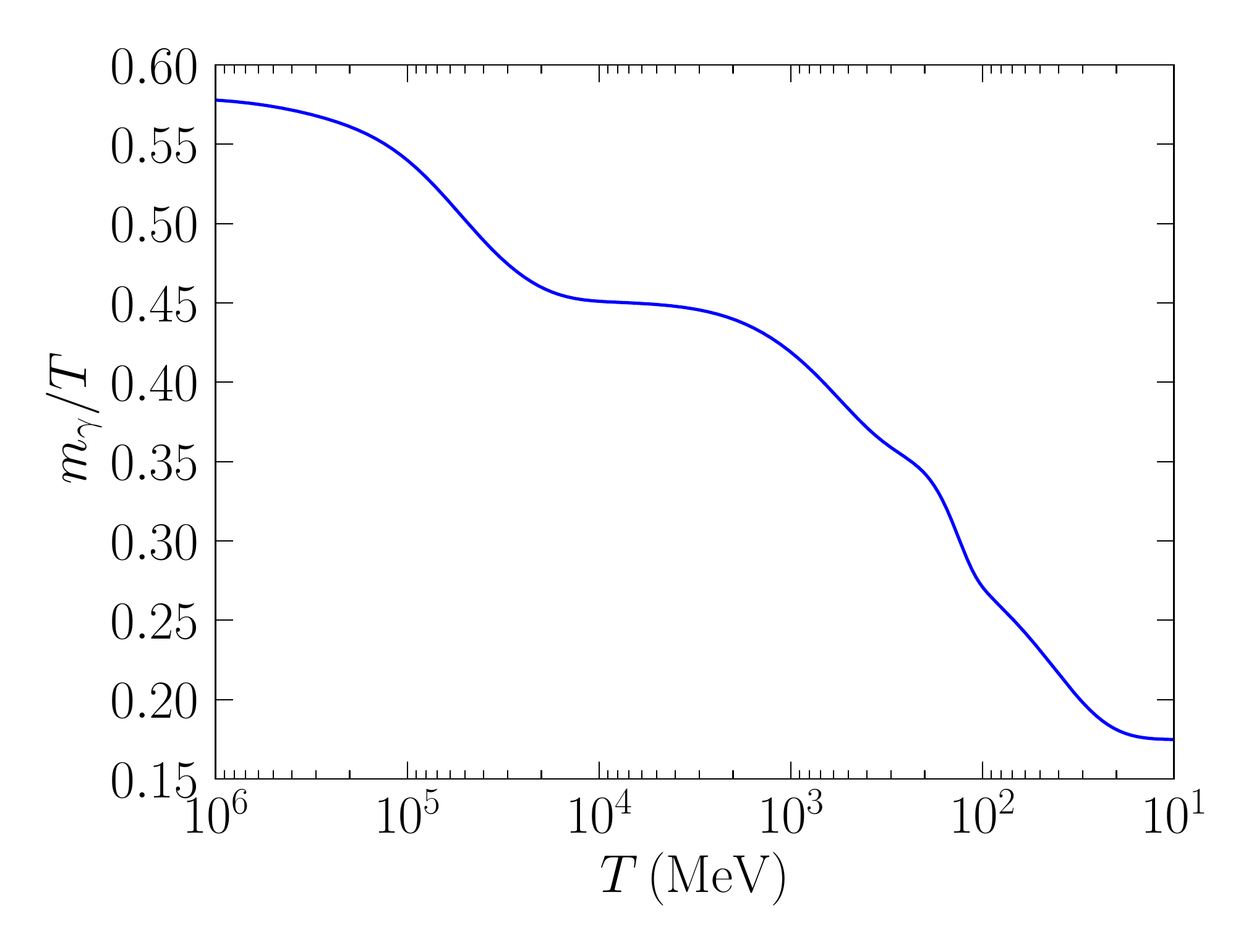}
  \caption{\label{fig:d_vs_t} Effective in-medium photon mass plotted as a function of plasma temperature [see Eq.\ \eqref{debye2}].
  \label{first}
  }
\end{figure}

The constraints on neutrino magnetic moment, such as those listed in the previous paragraph, are obtained considering neutrino electromagnetic scattering takes place in a plasma consisting of charged particles and antiparticles in the early universe. In such an environment screening of photons needs to be taken into account. We adopt a static screening prescription. Hence photons 
acquire an effective mass, which we denote by $m_{\gamma}$. The inverse of this mass is the Debye screening length for electromagnetic interactions. It is given by 
\begin{equation}
\label{debye2}
m_{\gamma}^2 = \frac{1}{\lambda_D^2} = 4\pi\alpha \sum_i q_i^2 \frac{\partial}{\partial \mu_i} [ n_i^{(-)} - n_i^{(+)}]. 
\end{equation} 
In Eq.\ \eqref{debye2}, $\alpha$ is the fine structure constant, $n_i^{(\mp)}$ is the number density for particles (antiparticles), respectively.  The partial derivative is with respect to the particle chemical potential and $q_i$ is the charge-coefficient of the particle for each particle-antiparticle pair (e.g., for an electron-positron plasma $q_i^2 = 1$).
Assuming thermal equilibrium and vanishing chemical potentials and masses for all particles, we obtain 
\begin{equation}
\label{debye}
m_{\gamma}^2\biggr\vert_{m_i=0,\mu_i=0} =  \frac{2\pi\alpha}{3}T^2 \sum_i q_i^2g_i,
\end{equation}
where $T$ is the plasma temperature and $g_i$ are the internal degrees of freedom from spin, color, etc.
The effective photon mass is plotted in Fig. \ref{first} as a function of the temperature. At very early times many particles are present in the 
plasma. As the universe evolves, the particle-antiparticle pairs annihilate one by one into lighter particles and no longer contributing to the effective photon mass in Eq.\ \eqref{debye2}. 

One can also explore dynamic effects on neutrinos coming from the plasma background \cite{1997NuPhB.503....3E,2000NuPhB.564..204A}. The authors of Refs. \cite{Carenza:2022ngg,Li:2022dkc} explored such effects and found changes in $N_{\rm eff}$ comparable to what we describe later in this article. This is perhaps not unexpected. For example in Ref. \cite{Hwang:2021kno} it was reported that light element abundances in BBN does not change when one considers dynamic screening. 

In our analysis we will use only the magnetic-moment term of the neutrino electromagnetic vertex function in Eq.\ \eqref{eq:form_factor}, namely $f_M(k^2)i\sigma_{\alpha\beta}k^\beta$.
We adopt the symbol $\kappa$ for the neutrino magnetic moment, defined as the magnetic dipole form factor in the forward-scattering limit, i.e., $\kappa\equiv f_M(k^2=0)$, which we scale to the Bohr magneton as
\beq\label{eq:kappa}
  \kappa = \mu\frac{e}{2m_e},
\eeq
using the dimensionless parameter $\mu$ and not to be confused with the chemical potentials in Eq.\ \eqref{debye2}.
We will include the effective photon mass from Eq.\ \eqref{debye2} when calculating scattering amplitudes and the resultant cross sections and rates.  For example: including in-medium effects modifies the well-known differential cross-section expression for elastic scattering from charged fermions to the following
\begin{equation}
\label{nmmxc}
\left(\frac{d \sigma}{dt}\right)_{\nu f}=
\frac{\pi q_f^2\alpha^2}{m_e^2} 
\mu^2 \frac{t}{(t-m_{\gamma}^2)^2} \frac{s+t-m_f^2}{s-m_f^2} 
\end{equation}
for each charged fermion with a mass $m_f$ and charge-coefficient $q_f$.
In Eq.\ \eqref{nmmxc}, the magnetic moment $\mu$ is given in units of Bohr magneton, hence $m_e$ in the prefactor is the same for all fermions since it comes from the definition of $\mu_B$.  $s$ and $t$ are the usual Mandelstam variables. 

We list the differential cross sections used in this work in Appendix \ref{app:a}. To obtain integrated cross sections, these expressions need to be integrated from $t_{\rm min} = -(s-m_i^2)^2/s$ to $t_{\rm max} =0$ to give the cross section as a function of $s$ and $T$, for each target mass $m_i$. Returning to the example for the differential cross section in Eq. \eqref{nmmxc}, integrating over $t$ gives 
\begin{eqnarray}
\sigma_{\nu f}(s) &=& \frac{\pi q_f^2
\alpha^2}{m_e^2} \mu^2 \left[ \left( 1+ \frac{2m_{\gamma}^2}{s-m_f^2} \right) \log \left(1+ \frac{(s-m_f^2)^2}{sm_{\gamma}^2} \right) \right. \nonumber \\ 
&-& \left. \frac{s-m_f^2}{s} - 1 + \frac{m_{\gamma}^2m_f^2}{sm_{\gamma}^2+ (s-m_f^2)^2}  \right] .\label{eq:sigma1}
\end{eqnarray}
We have explicitly given the cross section for elastic scattering off of fermions in Eq.\ \eqref{eq:sigma1}, but for all the other cross sections we numerically integrate the differential cross sections. 
After obtaining a cross-section like the one in Eq.\ \eqref{eq:sigma1}, we can calculate the thermal average of cross section multiplied by the Moller speed \cite{1991NuPhB.360..145G}, namely
\begin{equation}
  \langle\sigma_k v_{\rm Mol}\rangle = \dfrac{\dfrac{g_1g_2}{(2\pi)^6}\int d^3p_1\dfrac{1}{e^{E_1/T}+1}\int d^3p_2\,\sigma_k v_{\rm Mol}\dfrac{1}{e^{E_2/T}\pm1}}
  {\dfrac{g_1g_2}{(2\pi)^6}\int d^3p_1\dfrac{1}{e^{E_1/T}+1}\int d^3p_2\dfrac{1}{e^{E_2/T}\pm1}},
  \label{eq:sv_first}
\end{equation}
where the $k$ subscript indicates a specific scattering target and process.
In writing Eq.\ \eqref{eq:sv_first}, we have assumed equilibrium distributions for the incoming neutrino (labeled as particle 1) and the scattering target (particle 2) and ignored the Pauli blocking/Bose enhancement of the products.
The $\pm1$ in the distribution function for particle 2 corresponds to either fermions $(+)$ or bosons $(-)$.
For elastic scattering, the target particle is charged, whereas for annihilation it is an antineutrino.
Appendix \ref{app:b} gives simplified expressions for Eq.\ \eqref{eq:sv_first} in the case of scattering off of either fermions or bosons, and the annihilation process.

The last step in the procedure is to calculate the scattering rate using the thermally-averaged $\langle\sigma_k v_{\rm Mol}\rangle$number density of incoming neutrinos, $n_\nu$
\begin{equation}\label{eq:gamma_k}
  \Gamma_k = n_\nu\langle\sigma_k v_{\rm Mol}\rangle,
\end{equation}
Including the example specifically given in Eq.\ \eqref{nmmxc},
we calculate the individual rates for the processes of elastic scattering and annihilation for each charged particle target or product.  Summing over all of the individual rates gives us a total scattering rate $\Gamma_\nu$ as a function solely of temperature and $\mu$.

Returning again to the example in Eq.\
\eqref{eq:sigma1}, one can see that to leading order $\sigma$ does not scale with temperature.  In fact, $\langle\sigma v_{\rm Mol}\rangle$ also does not scale with $T$ to leading order.  Only the number density $n_\nu$ in Eq.\ \eqref{eq:gamma_k} provides a nontrivial $T^3$ scaling.  The Hubble expansion rate scales as $T^2$, implying that the magnetic moment interaction keeps the inactive neutrino states thermally populated at high temperatures, but become ineffective at lower temperatures.

\section{Cosmology}
\label{sec:cosmo}

The scattering and annihilation rates via the magnetic-moment vertex all scale as $\mu^2$, implying that increasing $\mu$ will increase the interaction rate and postpone the point when the inactive states decouple from the plasma.
In principle, decoupling could occur at low temperatures when the matter energy density comprises a significant fraction of the total energy density.
As we will show in Sec.\ \ref{sec:early}, current cosmological bounds imply that inactive neutrinos must decouple at early times, when the universe is dominated by radiation.

\begin{figure}
  \includegraphics[width=0.5\columnwidth]{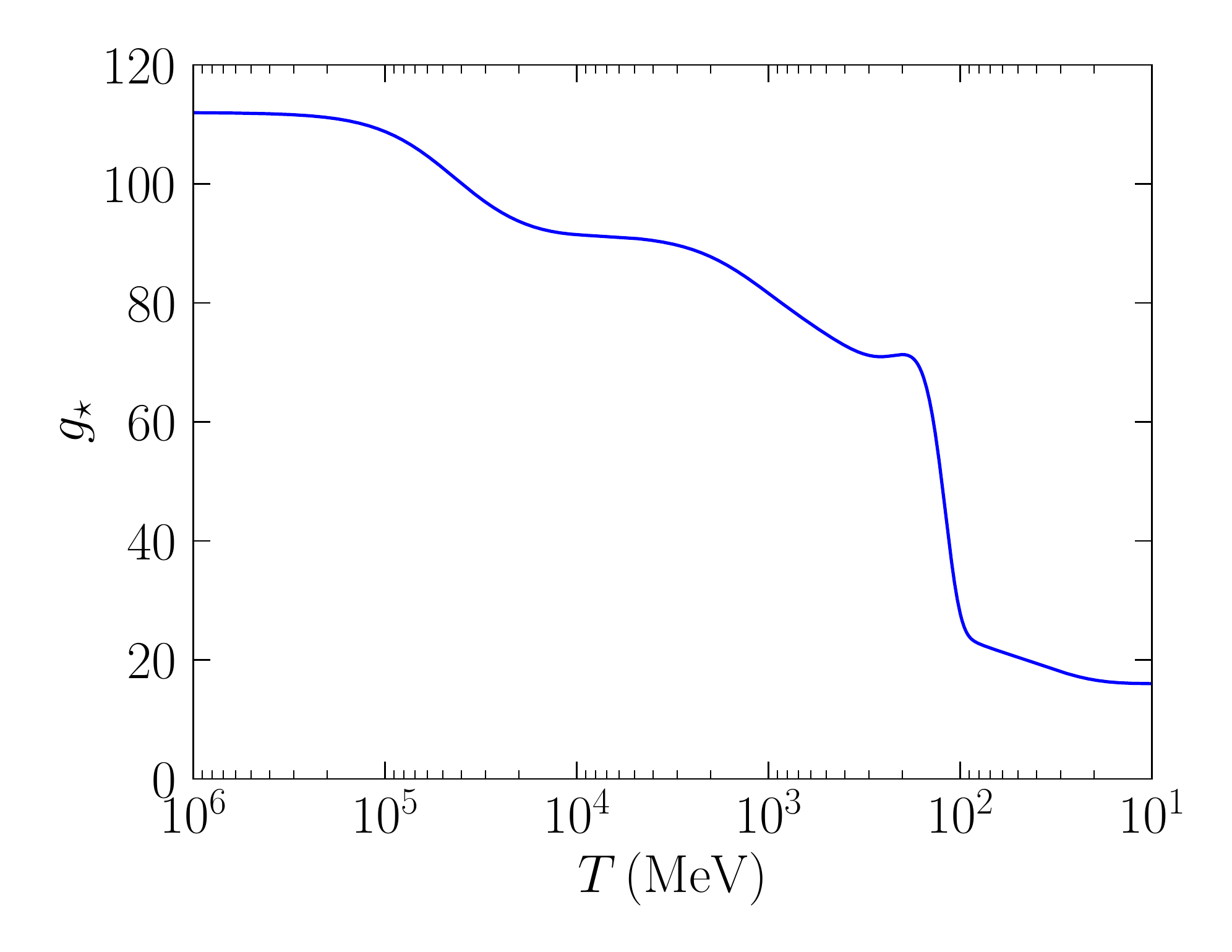}
  \caption{\label{fig:gstar_vs_t} Hubble expansion rate parameter \gstar
  plotted as a function of temperature.
  Included in the calculation of \gstar are the six degrees of freedom from the inactive neutrino states.
  }
\end{figure}

For radiation-dominated conditions, we will parameterize the energy density in two different ways.  When doing a calculation to determine decoupling, we use the parameter \gstar as an effective spin statistic constant \cite{1990eaun.book.....K}
\beq\label{eq:gstar}
  \rho = \frac{\pi^2}{30}\gstar T^4.
\eeq
When showing results on extra radiation energy density, we use the effective number of degrees of freedom, \neff, to parameterize the radiation energy density.  We delay discussion of \neff until Sec.\ \ref{sec:early}.
\gstar contains contributions from massless and massive particles.  To determine \gstar, we first calculate the total energy density using the appropriate Bose-Einstein or Fermi-Dirac (FD) equilibrium distribution function
\beq\label{eq:rho_stat}
  \rho = \sum\limits_i g_i\int\frac{d^3p}{(2\pi)^3}Ef_i(E),
\eeq
where the energy $E$ is related to the rest mass $m_i$ through $E=\sqrt{p^2+m_i^2}$.
We equate Eqs.\ \eqref{eq:gstar} and \eqref{eq:rho_stat} and solve for \gstar as a function of temperature.  Figure \ref{fig:gstar_vs_t} shows the relation between \gstar and plasma temperature employed in our decoupling calculations.  At the TeV scale, the entire standard model is present with ultra-relativistic kinematics.  As the universe expands and the temperature decreases, the equilibrium abundances of massive particles become Boltzmann suppressed and their respective degrees of freedom vanish.  The ``plateau-hill'' pattern in Fig.\ \ref{fig:gstar_vs_t} shows multiple instances of vanishing degrees of freedom.  When the temperature reaches $10\,{\rm MeV}$, only photons, electrons, and neutrinos contribute to \gstar.  
Included in the calculation of \gstar are the six inactive neutrino states at all temperatures.

To construct the plot in Fig.\ \ref{fig:gstar_vs_t} we make a number of simplifying assumptions.  First, we take the masses of the Higgs, vector bosons, and the top quark to be constant and equal to their vacuum values at all times. In reality, these and other particles acquire their masses during the EWT which occurs at $\sim 140\,{\rm GeV}$ \cite{2020JHEP...09..179R}.  As a result, our values for \gstar for $T\sim 140\,{\rm GeV}$ are an underestimate in Fig.\ \ref{fig:gstar_vs_t}.
Second, we use a fitting function to model the dynamics of the QHT centered at $T\sim 170\,{\rm MeV}$.  The fitting procedure produces a local maximum at $T\sim200\,{\rm MeV}$ when bound hadronic states coexist with free quarks and gluons.  Despite the local maximum, the energy density monotonically decreases with decreasing temperature at all times during the transition.
Appendix \ref{app:qht} gives details on the fitting procedure adopted from Ref.\ \cite{2014PhRvC..90b4915A}.

\begin{figure}
  \includegraphics[width=0.5\columnwidth]{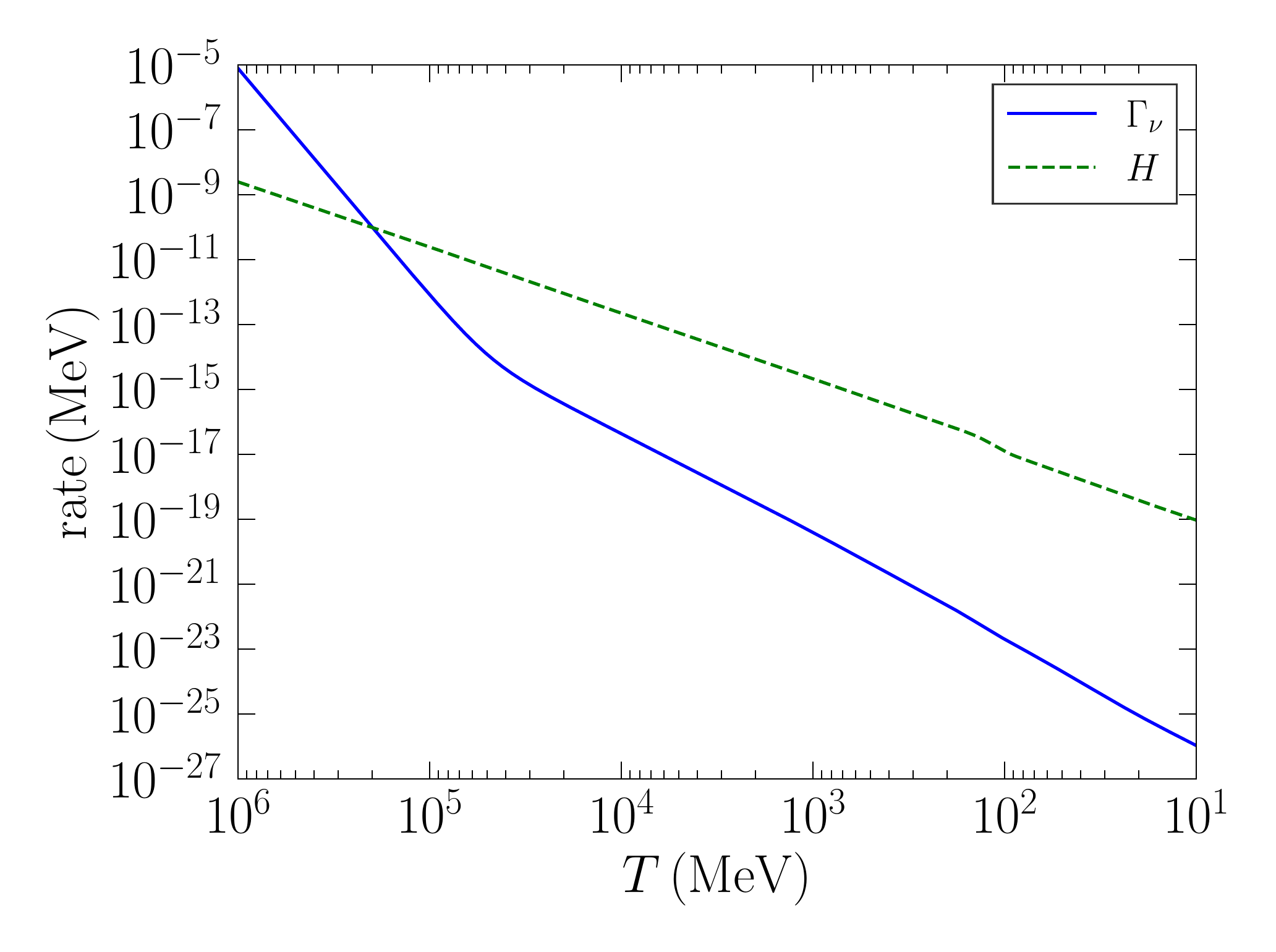}
  \caption{\label{fig:rates} Rates plotted against temperature.  The inactive
  neutrino scattering rate (solid blue) is for a magnetic moment strength
  $\mu=10^{-13}$.  Also given is the Hubble expansion rate (dashed green).
  }
\end{figure}

Equation \eqref{eq:gstar} ignores any contribution to the total energy density from matter or vacuum, appropriate for our purposes of inactive neutrino decoupling in the radiation-dominated regime.
We can calculate the Hubble expansion rate with \gstar to yield
\beq\label{eq:hub}
  H=\sqrt{\frac{8\pi}{3m_{\rm pl}^2}\rho}=\sqrt{\frac{4\pi^3}{45}\gstar}\,\frac{T^2}{\mpl}
\eeq
where $\mpl=1.2\times10^{19}\,{\rm GeV}$ is the Planck mass.  Equation \eqref{eq:hub} shows that the Hubble expansion rate scales as $T^2$.  The previous section showed that the magnetic-moment interaction rates scale as $\sim T^3$.
As a result, inactive neutrinos will maintain thermal equilibrium with the plasma at high temperatures, and eventually freeze-out and free-stream at lower temperatures.
Figure \ref{fig:rates} shows the total magnetic-moment interaction rate (solid blue) and Hubble expansion rate (dashed green) each as a function of temperature.  For this particular example, the magnetic-moment strength is taken to be $\mu=10^{-13}$.  For our purposes, we approximate decoupling as an instantaneous event when the interaction rate falls below the Hubble expansion rate
\beq
  \Gamma_\nu<H\implies {\rm decoupled}.
\eeq
For the example in Fig.\ \ref{fig:rates} we estimate the decoupling temperature as $T_{\rm dec}\simeq200\,{\rm GeV}$.
The magnetic-moment interaction rate scales as $T^3$ at low temperatures.  Once $W^\pm$ bosons are present in the plasma ($T\sim100\,{\rm GeV}$), the interaction rate increases dramatically, scaling as $T^7$.  This change in the scaling law is present in Fig.\ \ref{fig:rates} at a temperature scale comparable to the $W^\pm$ rest mass.

\begin{figure}
  \includegraphics[width=0.5\columnwidth]{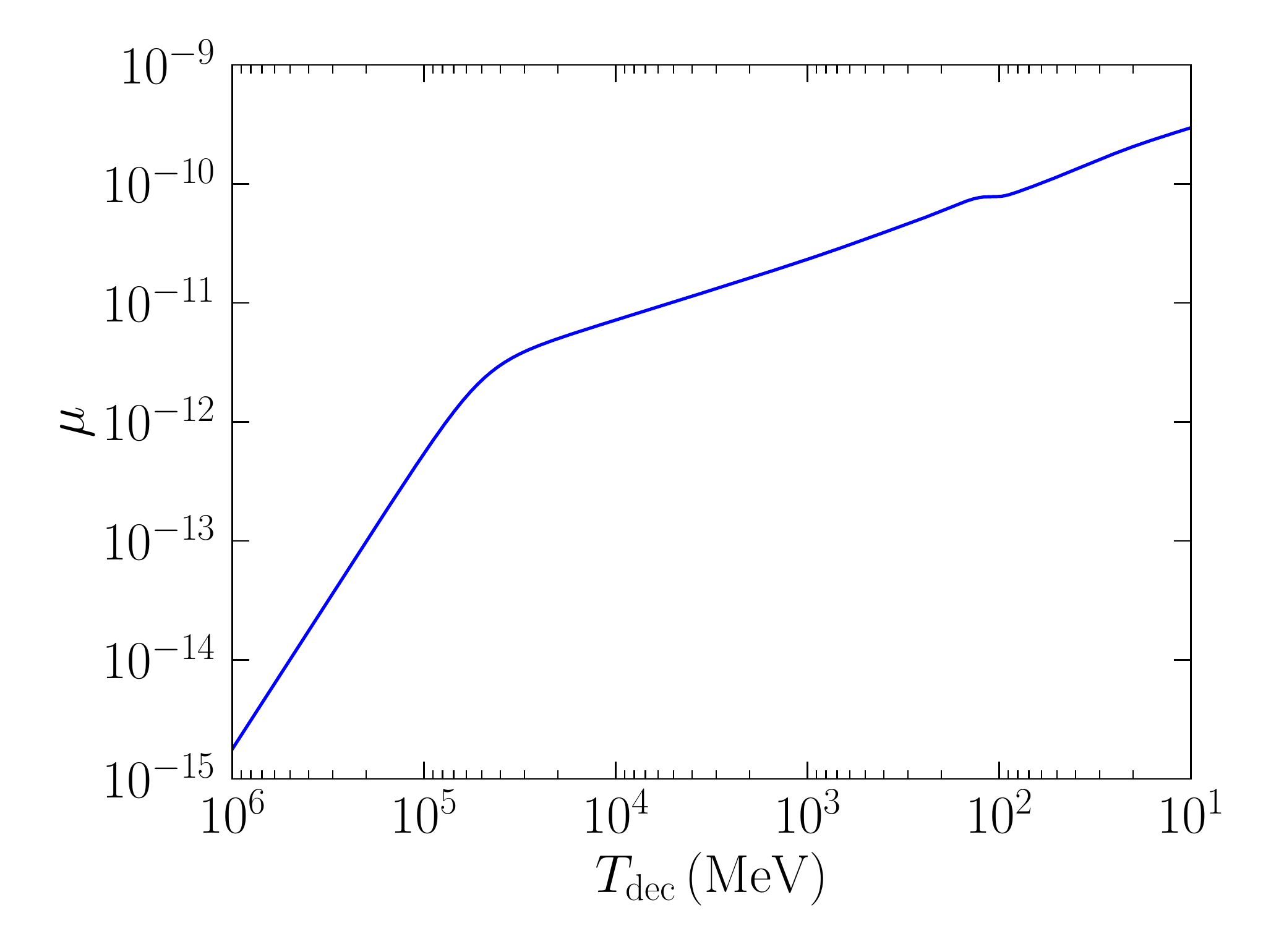}
  \caption{\label{fig:mu_vs_t} Magnetic moment strength $\mu$, corresponding to an interaction rate below the Hubble expansion rate, plotted as a
  function of decoupling temperature.
  }
\end{figure}

In our calculations, the magnetic-moment interaction rate is solely a function of the dynamical variable $T$ and the model parameter $\mu$.  All interaction rates are proportional to $\mu^2$, so an individual rate has the same temperature dependence as the blue curve in Fig.\ \ref{fig:rates} with an overall scaling dependent on $\mu$.
As a result, we can fix a decoupling temperature $T_{\rm dec}$ and solve for the corresponding $\mu$ by locating where the interaction rate falls below the Hubble expansion rate.
Figure \ref{fig:mu_vs_t} shows the magnetic moment strength as a function of the decoupling temperature.  The general behavior of the curve shows that increasing magnetic-moment strengths delays decoupling.
The shoulder at $T\sim100\,{\rm GeV}$ is again due to the presence of $W^\pm$ bosons in the plasma, akin to the behavior of the blue curve in Fig.\ \ref{fig:rates}.

\section{Early Universe Results}
\label{sec:early}

With the presence of the inactive Dirac states, there exists more energy density in the neutrino sector.  The inactive states have identical mass eigenvalues to those of the active neutrinos, and so their masses are small.  At early times, before photon decoupling, all of the neutrinos are ultrarelativistic and their energy density contributes to radiation.  At later times and the current epoch, the neutrino energy density contributes to matter.  In this section we discuss the implications at early times for the Cosmic Microwave Background (CMB) and BBN.

During atomic recombination, the radiation energy density is composed of photons, active neutrinos, and inactive neutrinos.  We assume that inactive-neutrino decoupling and active neutrino decoupling preserve the Fermi-Dirac spectra of the various neutrino species (see Refs.\ 
\cite{1997NuPhB.503..426D,2002PhLB..534....8M,2015NuPhB.890..481B, 2016PhRvD..93h3522G,2018PhR...754....1P,EscuderoAbenza:2020cmq,Akita:2020szl,Froustey:2020mcq,Bennett:2020zkv} among others on non-instantaneous decoupling).  The implication is that we can use temperature-like variables for the three components of the radiation.  Therefore, we write the radiation energy density during recombination as the following
\beq\label{eq:rhorad}
  \rhorad = \frac{\pi^2}{15}T^4 + 3\times\frac{7\pi^2}{120}T_a^4 + 3\times\frac{7\pi^2}{120}T_i^4,
\eeq
where $T_a$ is the active neutrino temperature-like quantity, and $T_i$ is a comparable quantity for the inactive neutrinos.  Conservation of comoving entropy gives the familiar relation between $T_a$ and $T$ \cite{1990eaun.book.....K}
\beq\label{eq:ratio_ap}
  \frac{T_a}{T} = \left(\frac{4}{11}\right)^{1/3}.
\eeq
The same principle applies for deducing the ratio $T_i/T$, and we find
\beq\label{eq:ratio_ip}
  \frac{T_i}{T} = \left(\frac{43}{11}\frac{1}{\gstarsdec}\right)^{1/3},
\eeq
where $\gstarsdec$ is the effective entropic degrees of freedom at inactive-neutrino decoupling $T_{\rm dec}$ (see Fig.\ \ref{fig:mu_vs_t}).
\gstarsdec is related to \gstardec by subtracting off the inactive neutrino degrees of freedom
\beq
  \gstarsdec = \gstardec - \frac{7}{8}\times6.
\eeq
Using the cosmological parameter \neff and the temperature ratios in Eq.\ \eqref{eq:rhorad}, we can relate \neff to \gstarsdec
\beq
  \neff = 3\left[1+\left(\frac{43}{11}\frac{1}{\gstarsdec}\right)^{4/3}\right].
\eeq

\begin{figure}
  \includegraphics[width=0.5\columnwidth]{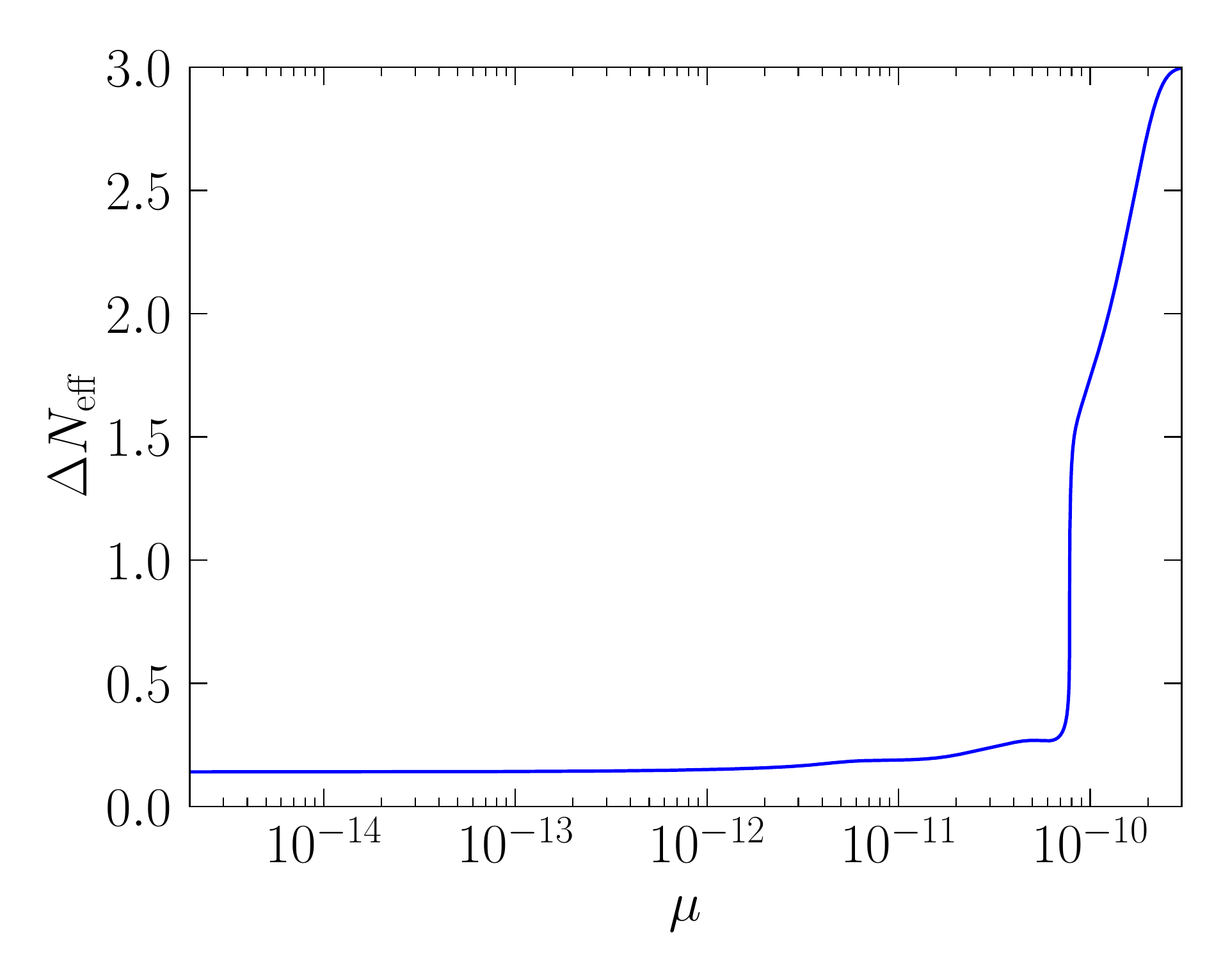}
  \includegraphics[width=0.5\columnwidth]{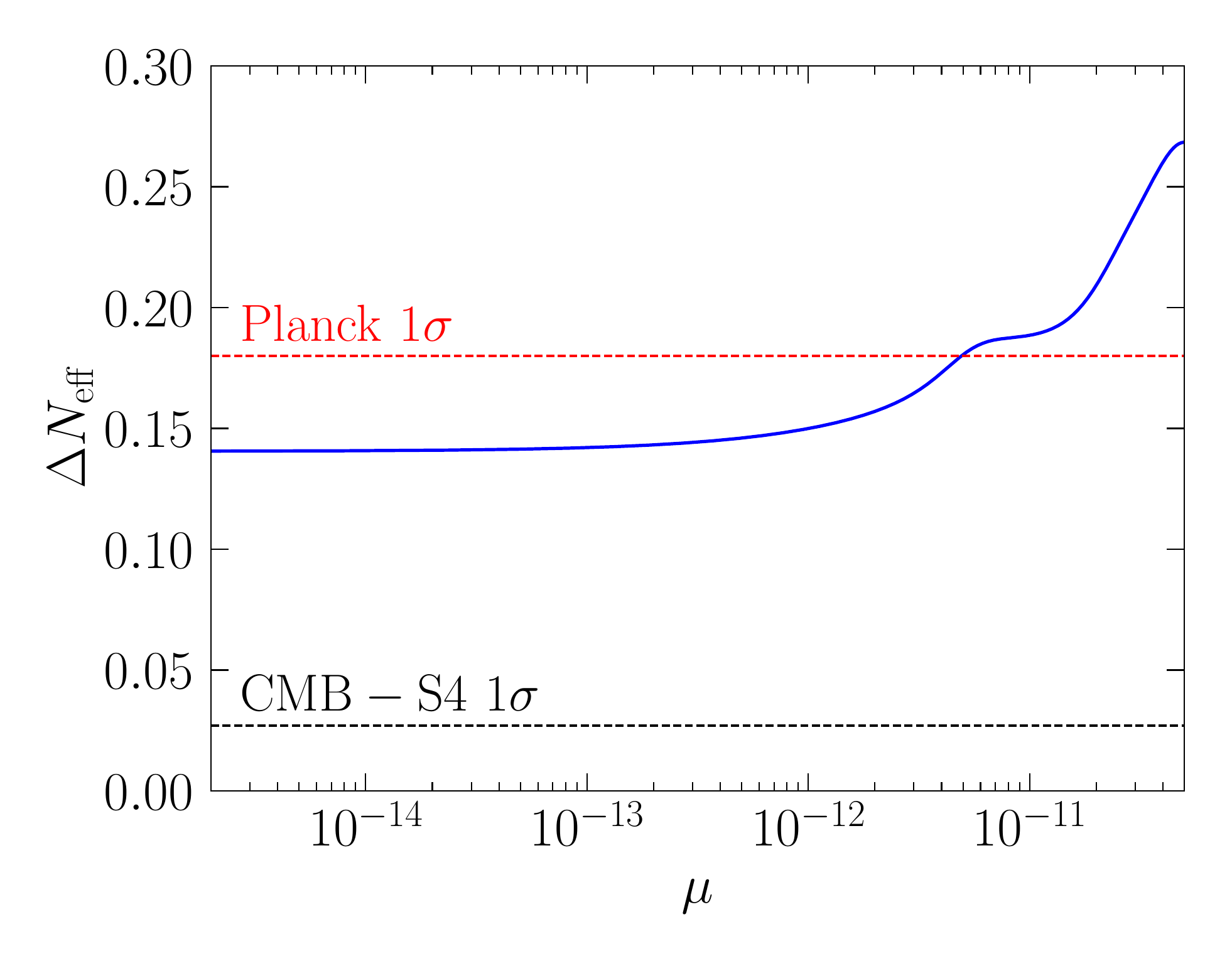}
  \caption{\label{fig:dneff_vs_mu} Change in \neff versus magnetic moment
  strength.  The top panel shows the entire range of magnetic moments explored in
  this work.  The bottom panel is a narrow range for low values of $\mu$.  Also
  plotted is the $1\sigma$ uncertainty from the Planck mission
  \cite{PlanckVI_2018} and the proposed $1\sigma$ uncertainty from CMB-S4
  \cite{cmbs4_science_book}.
  }
\end{figure}

Figure \ref{fig:dneff_vs_mu} shows the change in \neff in the presence of the inactive neutrino states.  The vertical axes are the change in \neff from 3, namely
\beq
  \Delta\neff\equiv\neff-3=3\left(\frac{43}{11}\frac{1}{\gstarsdec}\right)^{4/3}.
\eeq
The horizontal axes give a range of $\mu$.  In the top panel, we show the entire range of $\mu$ studied in this work, where the lower limit corresponds to $T_{\rm dec}\sim1\,{\rm TeV}$ and the upper limit to $T_{\rm dec}\sim1\,{\rm MeV}$.  For the large magnetic-moment strengths, the inactive neutrinos decouple at the same time as the active neutrinos do, implying that $T_a=T_i$ and both sectors contribute equally to \neff.

The bottom panel of Fig.\ \ref{fig:dneff_vs_mu} shows a restricted range of $\mu$, corresponding to decoupling before the QHT.  We have inserted horizontal lines to show the $1\sigma$ limits from the Planck mission \cite{PlanckVI_2018} and projections from CMB Stage IV \cite{cmbs4_science_book}.  We observe that at the level of $1\sigma$, $\mu\simeq5\times10^{-12}$ would produce a value of \neff in tension with Planck.  In the future, if CMB-S4 does not see any evidence of extra radiation energy density, then Dirac neutrinos could not have been in thermal equilibrium below the EWT to nearly $4\sigma$ level.

\begin{figure}
  \includegraphics[width=0.5\columnwidth]{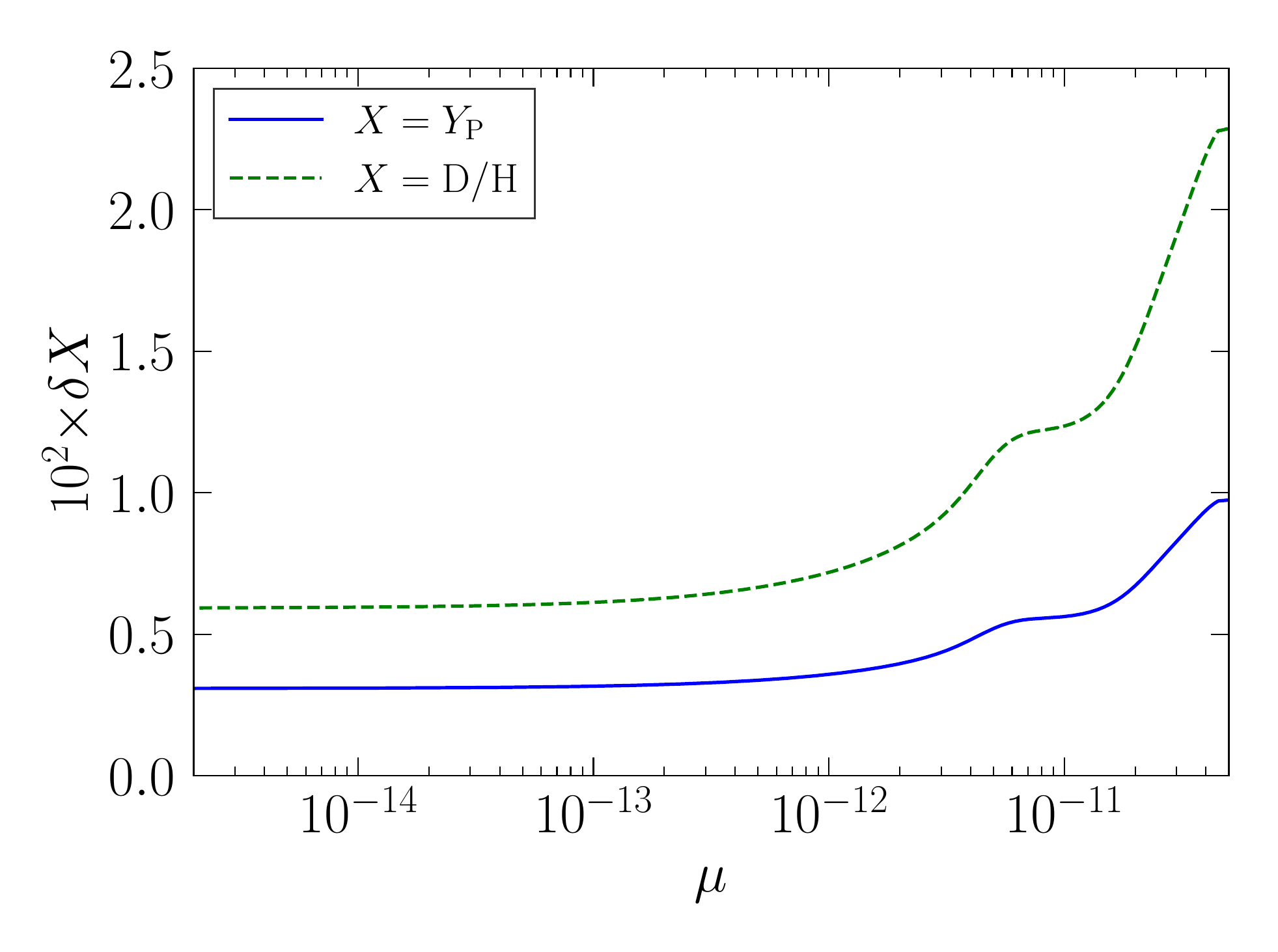}
  \caption{\label{fig:y_vs_mu}  Relative changes in primordial abundances
  plotted as a function of $\mu$.  The solid blue curve gives the relative change
  in the helium mass fraction $(\yp)$ and the dashed green curve gives the
  relative change in the deuterium abundance $({\rm D/H})$.
  }
\end{figure}

With an increase in radiation energy density, the Hubble expansion rate also increases which leads to an earlier epoch of weak freeze-out and nuclear freeze-out during BBN \cite{2023arXiv230112299G} [see Ref.\ \cite{2016RvMP...88a5004C} for the present status on BBN observations].  Figure \ref{fig:y_vs_mu} shows the relative differences in the helium-4 mass fraction, \yp, and the ratio of deuterium to hydrogen, \dtoh, as solid blue and dashed green lines, respectively.
The relative differences are computed by comparing to a baseline where there are no inactive Dirac states, and the active neutrinos decouple at a temperature of $10\, {\rm MeV}$.
The horizontal axis in Fig.\ \ref{fig:y_vs_mu} is the same range in $\mu$ as the bottom panel of Fig.\ \ref{fig:dneff_vs_mu}.  The shapes of the curves in Fig.\ \ref{fig:y_vs_mu} and $\Delta \neff$ in Fig.\ \ref{fig:dneff_vs_mu} are all simlar to one another as the abundances linearly scale with \neff in this range \cite{2004NJPh....6..117K}.  \dtoh is more sensitive to \neff and so has a larger deviation from the baseline value than \yp.  For $\delta(\dtoh)<1\%$, $\mu\lesssim4\times10^{-12}$, in line with the $1\sigma$ Planck limit from Fig.\ \ref{fig:dneff_vs_mu}.

\section{Later Universe Results}
\label{sec:later}

For the various epochs prior to photon decoupling, neutrino masses are small compared to the momenta, and so we approximated neutrinos as massless for calculations of energy density and interaction rates in Section \ref{sec:early}.
After neutrinos decouple from electrons and positrons (an epoch well before the photon-decoupling one), they continue to have ultra-relativistic kinematics and move at speeds nearly that of light.
During these later epochs neutrino kinematics will become increasingly non-relativistic in an expanding universe.
Accordingly, neutrino 3-momenta will redshift and asymptotically approach zero, implying the neutrino rest mass contribution evolves from a negligible to the dominant component of the neutrino energy density. As a result, we will need to discard our early-universe approximation of neutrinos being massless.

Typically, the dynamics of massive neutrinos is included in cosmology by extending the \lcdm model to include the ``sum of the light neutrino masses'' parameter, denoted as \summnu \cite{2006PhR...429..307L,Wong:2011ip}.
The presence of neutrino rest mass changes the growth of smaller versus larger-scale structure as neutrinos free stream during the initial stages of structure formation, but act as component of the total matter energy density at later stages.
The difference in the structure growth rates yields a modified matter power spectrum, which can be elucidated by considering weak gravitational-lensing of the CMB convolved with matter distributions from cosmological surveys \cite{2019BAAS...51c..64D}.
The transition between the small and large-scale regimes depends on \summnu, but also depends on the spectrum of the neutrinos.  If we introduce a non-thermal portion to the total cosmic neutrino spectrum via a low-energy contribution from the inactive Dirac states, we change the epoch when neutrinos become non-relativistic and hence the matter power spectrum and weak-lensing potential are appropriately altered. 

A theoretical calculation of the lensing potential in the presence of anomalous magnetic moments for massive neutrinos is beyond the scope of our exploratory work.
As an alternative to calculating the lensing potential, 
we will investigate the dependence of \summnu and $\mu$ on free-streaming.  Neutrinos move at speeds less than the speed of light, implying the more massive the neutrino the earlier it will become nonrelativistic over the history of the universe.  This implies a smaller free-streaming scale, $\lambda_{\rm fs}$.  We will use the free-streaming wavenumber $\kfs=2\pi a/\lambda_{\rm fs}$ (at scale factor $a$) to evaluate the roles of neutrino rest mass and anomalous magnetic moment strength on late-time cosmology.

Neutrinos will begin to move at speeds appreciably less than the speed of light once their momenta become comparable to their masses.  The results of \neff from the previous section showed that the inactive neutrinos must decouple from the plasma prior to the active neutrinos, implying that the comoving temperature quantity for the inactive states, $T_i$, is smaller than the counterpart quantity for the actives, $T_a$.  In fact, if $T_i<<T_a$, there is a possibility that the inactive states could become nonrelativistic in the early universe.  For the range of magnetic moment strength we consider in this work, the inactive states do not become non-relativistic until well after photon-decoupling.

\begin{figure}
  \includegraphics[width=0.5\columnwidth]{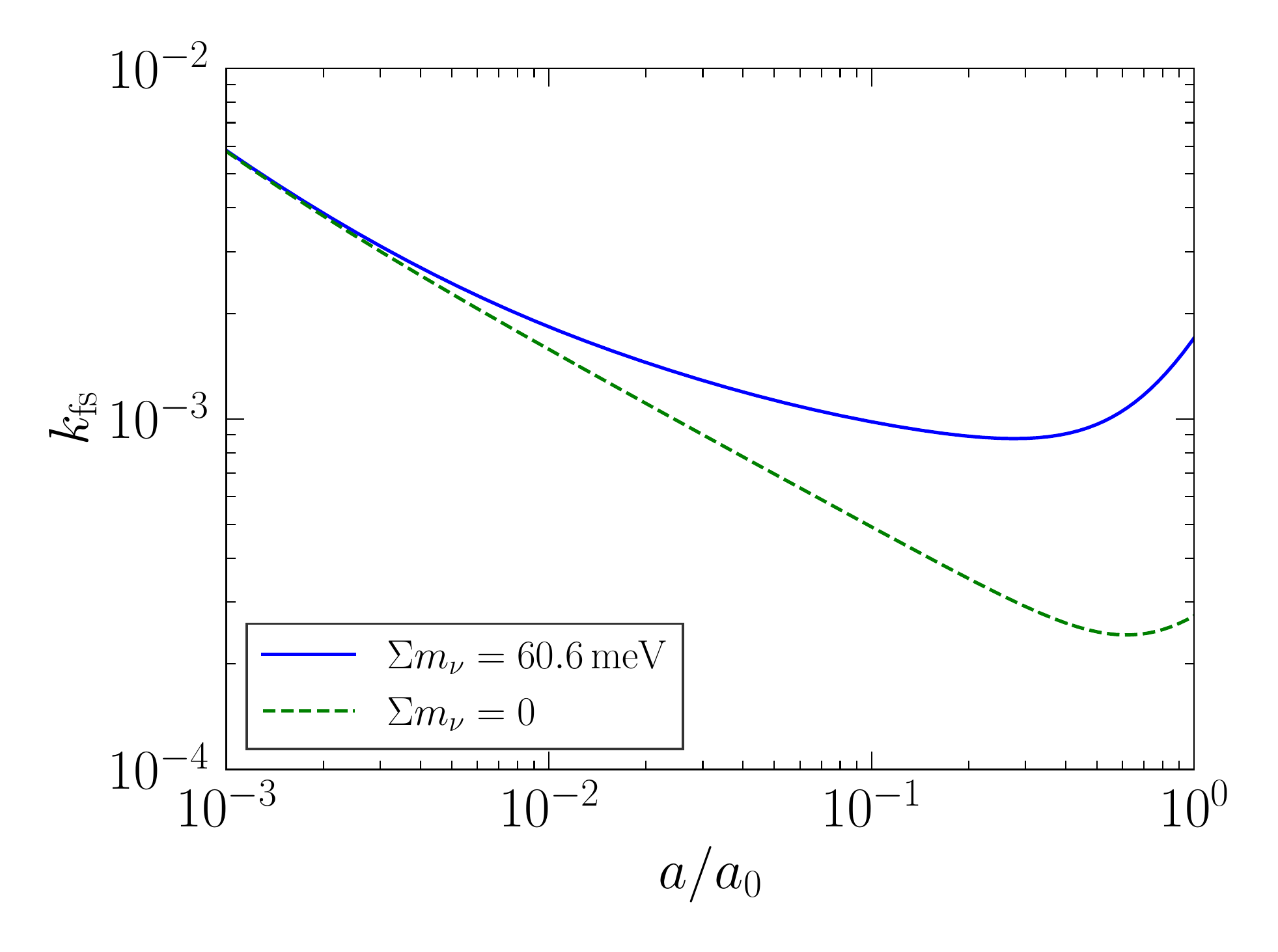}
  \caption{\label{fig:kfs_vs_a} \kfs plotted as a
  function of expansion parameter ratio $a/a_0$.  Solid blue line is for massive neutrinos as given in Eq.\ \eqref{eq:vth_av2} with $\summnu=60.6\,{\rm meV}$ in the normal mass ordering.  Dashed green line is \kfs if neutrinos were massless, i.e., $m_j=0$ in Eq.\ \eqref{eq:vth_av2}.  For both curves, we take the neutrino magnetic moment strength to be
  $\mu=1.88\times10^{-14}$. 
  }
\end{figure}

We adopt the definition of \kfs from Eq.\ (93) in Ref.\ \cite{2006PhR...429..307L}
\beq\label{eq:kfs}
  \kfs(t) = \sqrt{\frac{3}{2}}\frac{a(t)H(t)}{\vth(t)}
\eeq
where $t$ is the time coordinate and \vth is akin to the thermal speed with more explanation below.  For our purposes, we will use the scale factor $a$ as an independent variable.  If we consider the current epoch where $a=a_0$, we will denote the free-streaming wavenumber as \kfsz.  To incorporate the physics of anomalous magnetic moments, we will calculate the thermal speed using an ensemble average over the inactive and active states. We describe both the active and inactive neutrino states using FD distributions with their respective comoving temperature quantities.  All mass eigenstates for the active neutrinos have the same distribution with $T_a$, and similarly for the inactive states with $T_i$.  Due to $T_i\ne T_a$, the thermal speed is not a true thermal average, but instead an ensemble average.  Nevertheless, we adopt the nomenclature of thermal speed for consistency with the literature.  The thermal speed in our cosmology with anomalous magnetic moments is
\beq\label{eq:vth_av1}
  \vth =  \dfrac{2\times\sum\limits_{j=1}^{3}\int_0^\infty\frac{d^3p}{(2\pi)^3}
  \frac{1}{e^{p/T_a}+1}\frac{p}{E_j}+
  2\times\sum\limits_{j=1}^{3}\int_0^\infty\frac{d^3p}{(2\pi)^3}
  \frac{1}{e^{p/T_i}+1}\frac{p}{E_j}}
  {2\times\sum\limits_{j=1}^{3}
  \int_0^\infty\frac{d^3p}{(2\pi)^3}\frac{1}{e^{p/T_a}+1} +
  2\times\sum\limits_{j=1}^{3}
  \int_0^\infty\frac{d^3p}{(2\pi)^3}\frac{1}{e^{p/T_i}+1}},
\eeq
where $E_j=\sqrt{p^2+m_j^2}$ and the summations over $j$ are for the three separate mass eigenstates.   Note that $T_i$ and $T_a$ redshift with scale factor, so \vth also depends on scale factor.
At high temperatures -- equivalently low scale factor $a$ -- the neutrinos are ultrarelativistic and $p/E\sim1$ and so $\vth\sim1$.  Conversely, at high scale factor $p/E<1$.
We can simplify Eq.\ \eqref{eq:vth_av1}
to the following
\beq
  \vth =  \frac{2}{9\zeta(3)[T_a^3+T_i^3]}\int_0^\infty d\eps
  \frac{\eps^2}{e^{\eps}+1}
  \sum\limits_{j=1}^{3}\left[\dfrac{T_a^3}{
  \sqrt{1+\left(\frac{m_j}{\eps T_a}\right)^2}}
  +\dfrac{T_i^3}{
  \sqrt{1+\left(\frac{m_j}{\eps T_i}\right)^2}}\right],
  \label{eq:vth_av2}
\eeq
where we have used $\epsilon=p/T_x$ for either $x=a,i$.  Figure \ref{fig:kfs_vs_a} shows the evolution of \kfs as a function of the ratio $a/a_0$.  To plot \kfs, we need the following model input parameters: the ratios of temperature quantities $T_i/T$ and $T_a/T$; and the light neutrino mass eigenstates.
$T_i/T$ is a function of the magnetic moment strength $\mu$ implied in Eq.\ \eqref{eq:ratio_ip}, and $T_a/T=(4/11)^{1/3}$.  For the mass eigenstates, we use the parameter \summnu and specify an ordering, either normal or inverted, using the solar and atmospheric mass splitting values where appropriate \cite{Zyla:2020zbs}.
For both curves in Fig.\ \ref{fig:kfs_vs_a}, we pick $\mu=1.88\times10^{-14}$.  The solid blue curve uses $\summnu=60.6\,{\rm meV}$ with a normal mass ordering, i.e., a smallest mass eigenvalue $m_1=1\,{\rm meV}$.  To show the effect of mass on \kfs, we also plot a dashed green curve using massless neutrinos, i.e., using $\vth=1$.
The neutrino energy density differs between the two cosmologies.  For the purposes of comparing the two models, we preserve the Hubble expansion rate at the current epoch by adjusting the vacuum energy density, i.e., we decrease $\rho_{\Lambda}$ for increasing \summnu.
For $a/a_0\gtrsim10^{-3}$, we see that the blue curve diverges from the green curve due to massive neutrinos becoming nonrelativistic.  The increase in \kfs corresponds to a decrease in power on small scales at later times.
The divergence increases to the current epoch, at which point \kfs differs by an order of magnitude between the two cosmologies.  For concreteness, we give the two values at the current epoch
\begin{align}
  \kfsz(\summnu=60.6\,{\rm meV}) &= 1.70\times10^{-3},\label{eq:kfs_60}\\
  \kfsz(\summnu=\hphantom{60.}0\,{\rm meV}) &= 2.74\times10^{-4}.\label{eq:kfs_0}
\end{align}

We have employed a cosmology in Fig.\ \ref{fig:kfs_vs_a} where the magnetic-moment interaction populates the inactive states resulting in a larger value of \neff.
At this point, we give a brief digression to discuss how neutrino rest mass affects \kfs when magnetic moments are not present but $\Delta\neff>0$.
For a cosmology where $\Delta\neff=0$ yet neutrinos have non-zero masses,
we can compensate for the larger neutrino energy density by using a smaller vacuum energy density fraction, $\Omega_\Lambda$, to preserve the Hubble expansion rate at the current epoch.  
We use the same prescription when $\Delta\neff>0$, regardless of whether that extra radiation energy density is from neutrinos or some other undetermined particles.
For the base values of $\Omega_\Lambda$ and the cold dark matter fraction we use in this work \cite{PlanckVI_2018}, a small decrease in $\Omega_\Lambda$ implies a younger universe, and therefore a lower value of the free streaming length and larger value of \kfs.
When we introduce the inactive Dirac states via $\mu=1.88\times10^{-14}$, \kfsz does not vary at all from the base cosmology if neutrinos were massless.  This is a result of $\vth=1$ and our prescription of fixing the Hubble expansion rate at the current epoch to be the same in Eq.\ \eqref{eq:kfs} regardless of the cosmological model.  On the other hand, for massive neutrinos with $\summnu=60.6\,{\rm meV}$ and $\vth\ne1$, the value in Eq.\ \eqref{eq:kfs_60} is 5\% higher than the comparable cosmological model with massive neutrinos and unpopulated inactive states.

\begin{figure}
  \includegraphics[width=0.5\columnwidth]{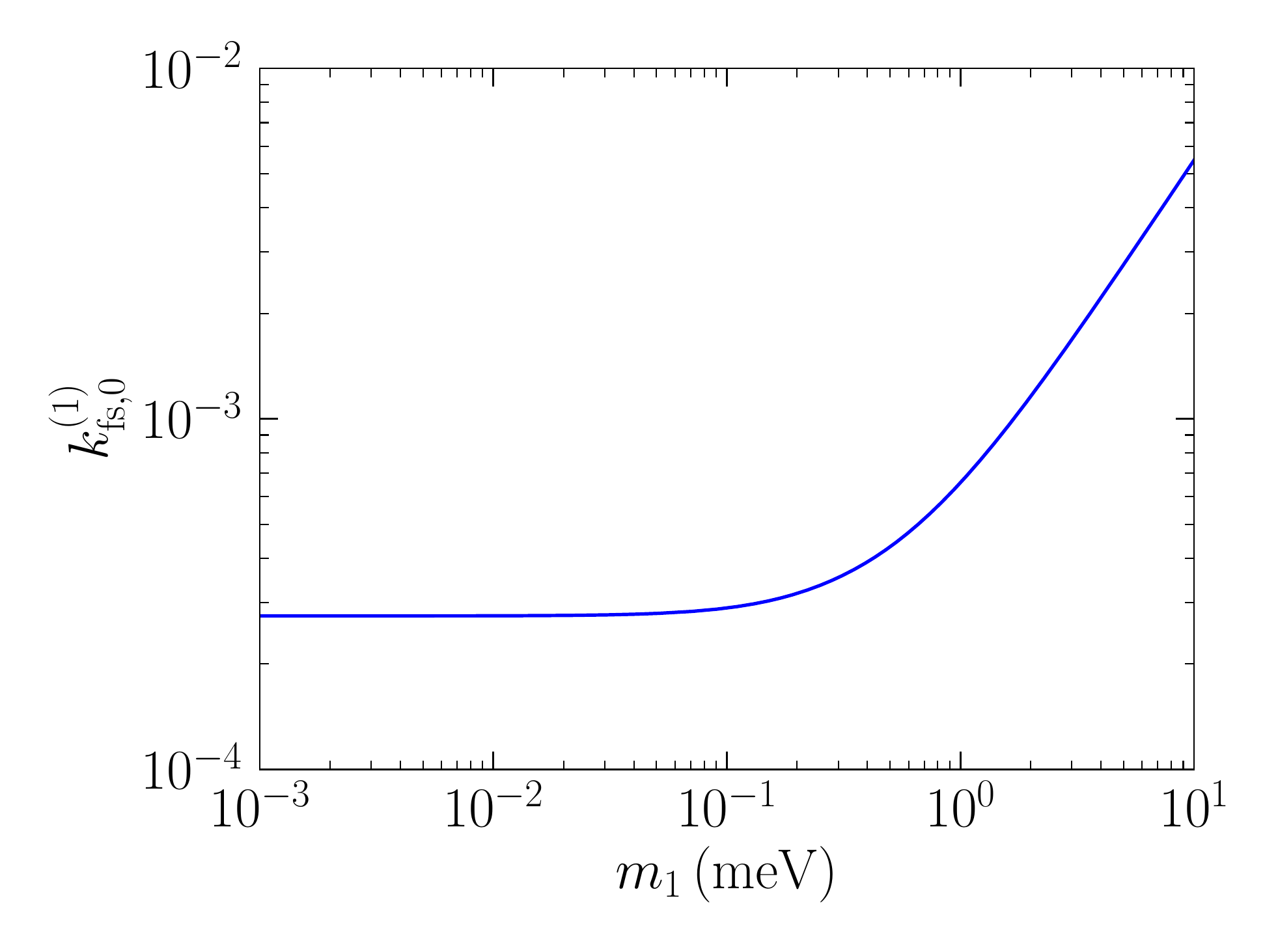}
  \caption{\label{fig:kfs_vs_m1} Free-streaming wavenumber for the neutrinos with lightest mass eigenvalue $m_1$ plotted as a function of $m_1$ (meV).  $k_{\rm fs, 0}^{(1)}$ is calculated by setting the summations in Eq.\ \eqref{eq:vth_av1} for \vth only to range over $j=1$, and includes active and inactive neutrinos for the model
  $\mu=1.88\times10^{-14}$.
  }
\end{figure}

Figure \ref{fig:kfs_vs_a} shows the impact on \kfs when neutrinos become nonrelativistic.
At the current epoch, the CMB photon temperature is $T=2.726\,{\rm K}$, implying that the active neutrino temperature is $T_a=0.17\,{\rm meV}$.  For $\summnu=60.6\,{\rm meV}$ in a normal ordering, the neutrinos with the two heavier mass eigenvalues are nonrelativistic and have been for much of the history of the universe.  Decreasing \summnu below 60.6 meV to its absolute minimum of 59.6 meV only slightly changes the values of $m_2$ and $m_3$, but has a significant effect on \kfs.  Although \summnu changes by less than $2\%$, \kfs decreases by nearly $60\%$.  The decrease is entirely due to the kinematics of the neutrinos with the smallest mass eigenvalue $m_1$.
Figure \ref{fig:kfs_vs_m1} shows the quantity $k_{\rm fs, 0}^{(1)}$ plotted against $m_1$. $k_{\rm fs, 0}^{(1)}$ is the free-streaming wavenumber for only the neutrinos with $m=m_1$.  We calculate $k_{\rm fs, 0}^{(1)}$ by first replacing the summations in Eq.\ \eqref{eq:vth_av1} for \vth with single calculations where $m_j=m_1$, and then calculating the free-streaming wavenumber with Eq.\ \eqref{eq:kfs}.  For $m_1=1\,{\rm meV}$, Fig.\ \ref{fig:kfs_vs_m1} shows that the distribution of lightest mass neutrinos is in transition from ultrarelativistic to nonrelativistic at the current epoch.  Decreasing $m_1\lesssim0.1\,{\rm meV}$ ensures the lightest neutrinos are ultrarelativistic for the entire history of the universe.

We have used a model where $\mu=1.88\times10^{-14}$ when plotting $k_{\rm fs, 0}^{(1)}$ in Fig.\ \ref{fig:kfs_vs_m1}.  For other values of $\mu$, $k_{\rm fs, 0}^{(1)}$ vs.\ $m_1$ would look qualitatively identical to Fig.\ \ref{fig:kfs_vs_m1}.  One of the quantitative differences for differing $\mu$ models is the value of $m_1$ where the kinematics of the neutrinos transition from relativistic $(E=\sqrt{p^2+m^2})$ to ultrarelativistic $(E=p)$, represented by the ramp-up from the plateau of $k_{\rm fs, 0}^{(1)}$ in Fig.\ \ref{fig:kfs_vs_m1}.
There are two competing effects which alter the point of departure from the plateau.
First, the inactive neutrino temperature is always smaller than the active neutrino temperature, and so the inactive neutrinos with a given mass depart from ultrarelativistic kinematics before the actives with that same mass in the history of the universe.  Figure \ref{fig:mu_vs_t} implies that the ratio $T_i/T_a$ decreases with decreasing $\mu$, so models with smaller $\mu$ have earlier points of departure in Fig.\ \ref{fig:kfs_vs_m1}.
However, in opposition to this first effect is the fact that smaller $T_i$ implies a smaller number density for the inactive neutrinos.  A smaller number density increases \kfs (and by extension $k_{\rm fs, 0}^{(1)}$) in Eq.\ \eqref{eq:vth_av2}, implying models with smaller $\mu$ would have a later point of departure in Fig.\ \ref{fig:kfs_vs_m1}.

\begin{figure}
  \includegraphics[width=0.5\columnwidth]{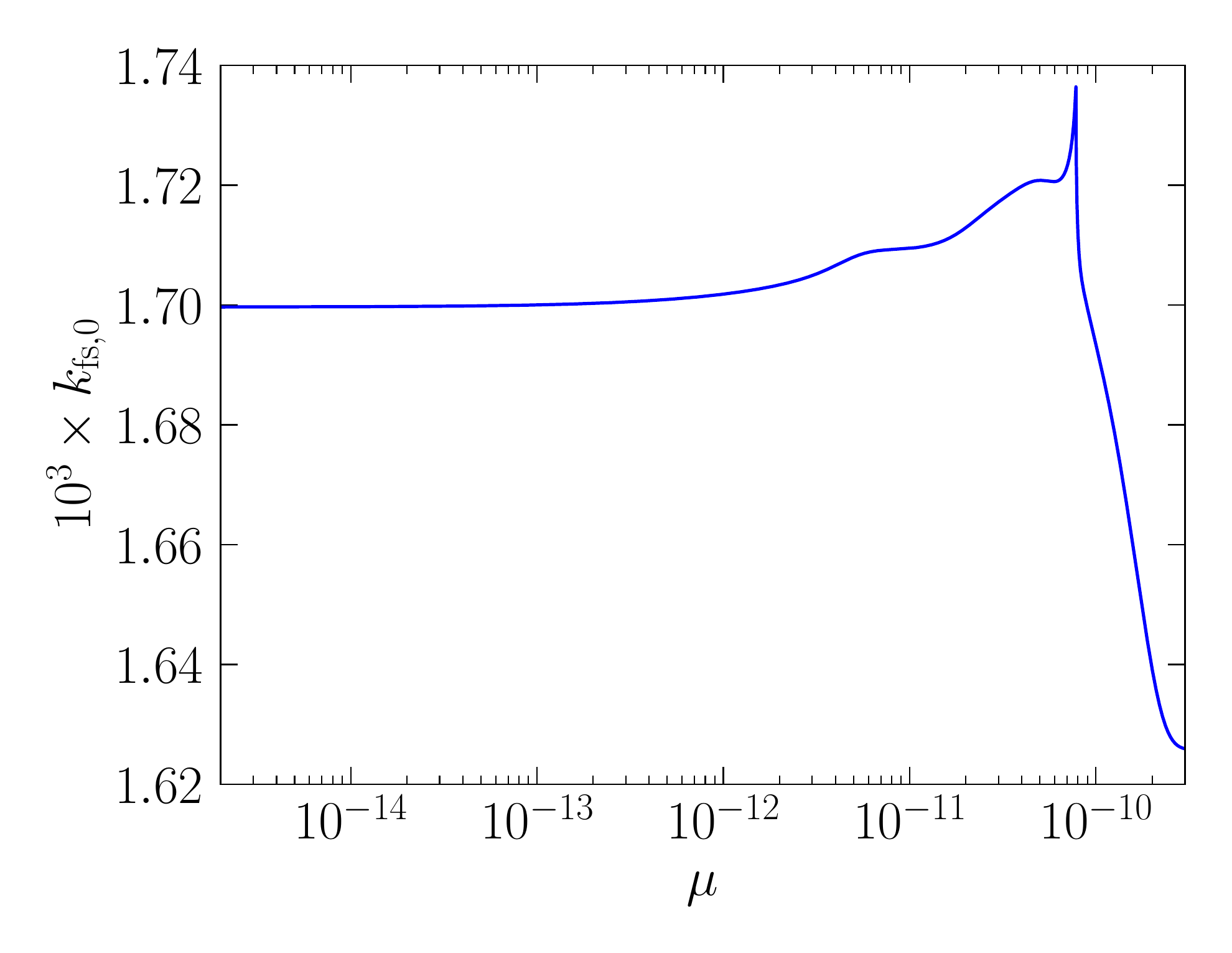}
  \caption{\label{fig:kfs_vs_mu}  \kfsz plotted
  as a function of magnetic moment strength $\mu$.  $\summnu=60.6\,{\rm meV}$ in the normal ordering. 
  }
\end{figure}

To show the competition between temperature and number density, we consider how \kfsz varies with $\mu$ for a fixed $m_1$ or equivalently a fixed \summnu.
Figure \ref{fig:kfs_vs_mu} shows how \kfsz changes with $\mu$ for $\summnu=60.6\,{\rm meV}$ in a normal ordering.  The higher values of $\mu$ show the effect of $T_i/T_a$ close to unity, whereas the lower values show the effects of a smaller number density.  There is a global maximum for these models at $\mu\lesssim10^{-10}$, corresponding to a decoupling temperature $T_{\rm dec}\simeq100\,{\rm MeV}$.  This epoch occurs in proximity to the QHT and therefore the exact value of the global maximum is dependent on the treatment of the QHT.  The shape of the curve in Fig.\ \ref{fig:kfs_vs_mu} is a function of the temperature ratio $T_i/T_a$.  A larger value of \summnu acts to shift the curve down to smaller values of \kfsz while preserving $T_i/T_a$ and the shape of that curve.

\begin{figure}
  \includegraphics[width=0.5\columnwidth]{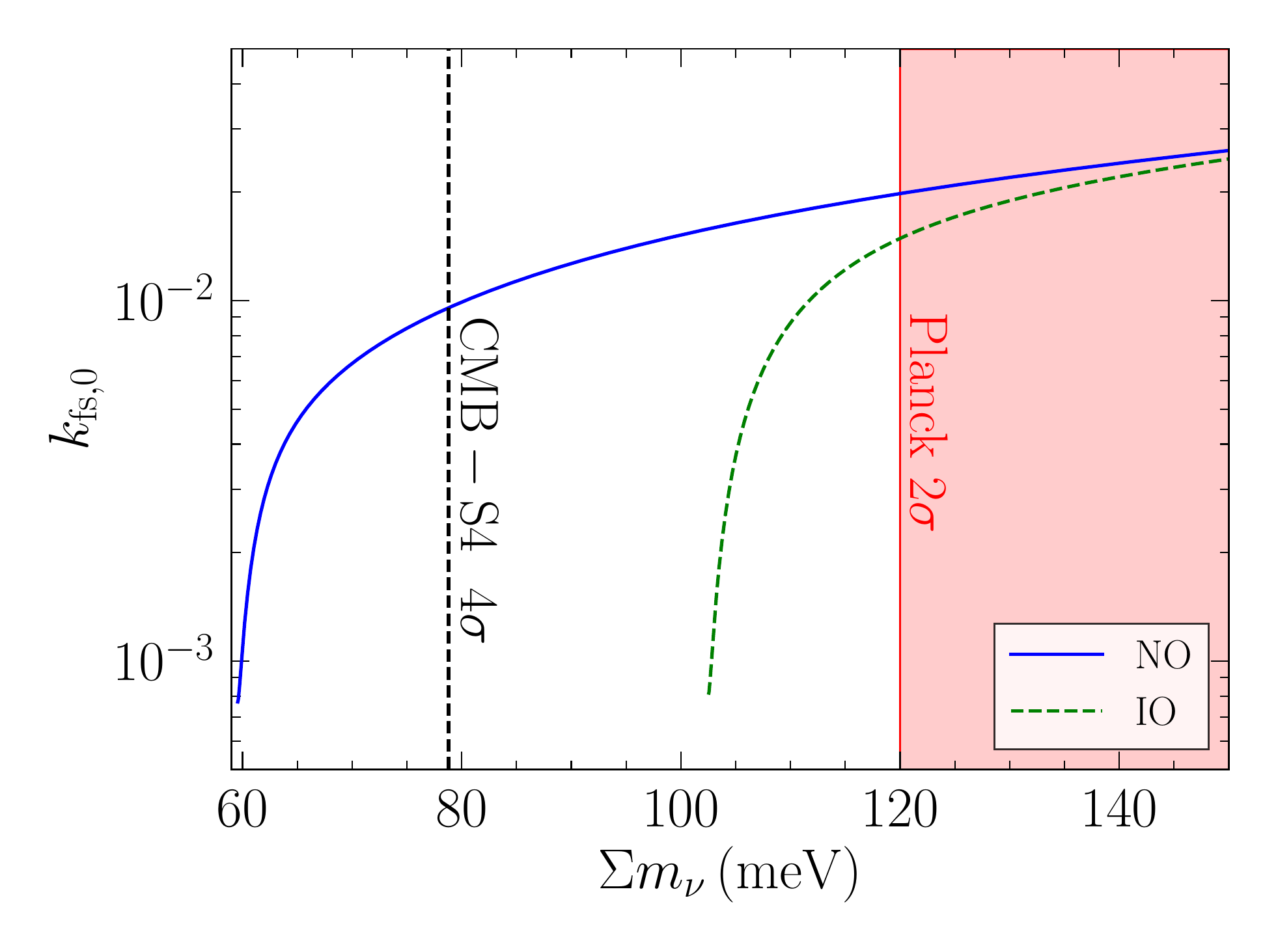}
  \caption{\label{fig:kfs_vs_summnu} \kfsz from Eq.\ \eqref{eq:vth_av2} plotted as a
  function of \summnu (meV).  Solid blue line is for the normal ordering, and dashed green line for the inverted ordering.  Plot is for the model
  $\mu=1.88\times10^{-14}$.
  Also plotted are the current constraints on \summnu from the Planck mission at $2\sigma$ (Ref.\ \cite{PlanckVI_2018}) and a forecast from CMB-S4 at $4\sigma$ (Ref.\ \cite{cmbs4_science_book}).
  }
\end{figure}

Finally, we show how \kfsz changes with \summnu for the normal (solid blue) and inverted (dashed green) orderings in Fig.\ \ref{fig:kfs_vs_summnu}.
Both curves use a model where $\mu=1.88\times10^{-14}$.  The apparent asymptote at the lowest values of \summnu for each ordering are a result of neutrinos with mass eigenvalue $m_1$ staying ultrarelativistic until the current epoch, analogous to the descent to the plateau in Fig.\ \ref{fig:kfs_vs_m1}.  \kfsz has a smaller minimum value for the normal ordering as a result of a smaller neutrino energy density and older universe. 
We have plotted the $2\sigma$ constraint on \summnu from the Planck mission \cite{PlanckVI_2018} and a $4\sigma$ forecast from CMB-S4 \cite{cmbs4_science_book}.  If CMB-S4 finds a nonzero value for \summnu, scales such as the free-streaming length would differ between the two orderings.

\section{Conclusions}
\label{sec:conclusions}

The mechanism which generates the neutrino mass has yet to be determined, and as a result the nature of whether neutrinos are Majorana or Dirac is unknown.  In this work, we have considered the cosmological implications on the existence of the inactive Dirac states and how they may be thermally populated at an early epoch in the history of the Universe.  
Both Dirac and Majorana neutrinos could impact \neff through undetected interactions. For example, electromagnetic scattering of electrons and positrons with Majorana neutrinos changes the spectra of the active component through a heat flow from the electromagnetic plasma into the neutrino seas \cite{Vassh_tdecoup}.  In the models considered in this work, electromagnetic scattering channels of charged particles with Dirac neutrinos populate the inactive states while preserving the FD spectra of the active states.

Motivated by the search for phenomenological differences between the Majorana versus Dirac nature of neutrinos, we studied a class of interactions between Dirac neutrinos and standard-model particles not mediated by the weak force.  As a hypothesis, the model we employed utilizes anomalous neutrino magnetic moments and an associated electromagnetic vertex, namely $i\kappa\sigma_{\alpha\beta}k^\beta$ in Eq.\ \eqref{eq:form_factor}, to couple neutrinos to charged leptons, quarks, and $W^{\pm}$ bosons.  We calculated thermally-averaged cross sections for both the elastic scattering and the annihilation processes using the Debye screening length to reflect the bath of charged particles present in the plasma of the Early Universe. 
In addition, we calculated the two scattering cross sections between neutrinos and $W^{\pm}$ for the first time using a magnetic moment vertex (see Appendix \ref{app:a}). With these cross sections, we compare the associated scattering rates to the Hubble expansion rate to find a decoupling temperature as a function of the neutrino magnetic moment, parameterized using $\mu$.  Figure \ref{fig:mu_vs_t} shows the relation between the decoupling temperature and $\mu$, where the scaling law changes around $T\sim100\,{\rm GeV}$ due to the presence of the $W^{\pm}$ bosons. 

With the additional neutrino states populated, we were able to calculate changes to \neff, the primordial abundances \yp and \dtoh, and the free-streaming wavenumber.
Our strongest limits on $\mu$ come from those of \neff using parameter estimations from the Planck mission \cite{PlanckVI_2018}.
Table \ref{tab:limits} gives limits from experiment \cite{Beda:2012zz,1993PhRvD..47...11A,2001PhRvD..63k2001A, DONUT:2001zvi,Borexino:2008dzn},
other astrophysical or cosmological sources \cite{Vassh_tdecoup,Capozzi:2020cbu,Mori:2020niw,Mori:2020qqd},
and finally our current work, and will be further constrained with upcoming CMB experiments. 
More specifically, the relation between \neff and $\mu$ in Fig.\ \ref{fig:dneff_vs_mu} is sensitive to how one treats the EWT and QHT.  In terms of computing direct energy density, the QHT is obviously the more sensitive probe.  However, if CMB-S4 pushes the limits on \neff to epochs preceding the QHT, the EWT becomes more pertinent.  We caution though that the EWT is not well-understood and the scaling relations between the rates and $\mu$ may differ above electroweak symmetry breaking.  Our results above $\sim100\,{\rm GeV}$ should be taken as extrapolations.

\begin{table}
    \centering
    \begin{tabular}{ccccc}
         Method & Limit (in units of $\mu_B$) & Notes \\
        \hline
        & & \\
        Reactor &  $2.9\times10^{-11}$ &  GEMMA \cite{Beda:2012zz}\\
        Accelerator $\nu_e$-$e^-$ & $10^{-11}$  & LAMPF \cite{1993PhRvD..47...11A}\\
        Accelerator $(\nu_\mu,\overline{\nu}_{\mu})$-$e^-$ & $6.8\times10^{-10}$ & LSND \cite{2001PhRvD..63k2001A}\\
        Accelerator $(\nu_\tau,\overline{\nu}_{\tau})$-$e^-$ &  $3.9\times10^{-7}$  &  DONUT \cite{DONUT:2001zvi} \\
        Dark Matter Direct Detection & 
        $4.9 \times 10^{-11}$ & PandaX \cite{PandaX-II:2020udv} \\
        Dark Matter Direct Detection ($\nu_e$) & $0.9 \times 10^{-11}$ & XENONnT \cite{2023PhLB..83937742K}
        \vspace{10 pt}\\
        Solar $({}^7{\rm Be})$ & $5.4 \times 10^{-11}$ & Borexino \cite{Borexino:2008dzn}\\
        Red Giant Stars & $1.2 \times 10^{-12}$ & Ref. \cite{Capozzi:2020cbu} \\
        Cepheid Stars &  $2 \times 10^{-10} $       & Ref. \cite{Mori:2020niw} \\
        Lithium in red clump stars & $10^{-12}$     & Ref. \cite{Mori:2020qqd}
        \vspace{10 pt}\\
        Cosmology (Majorana) & $10^{-10}$ & Ref. \cite{Vassh_tdecoup}\\
        Cosmology (Dirac) & $5\times10^{-12}$ & This work (\neff limit from Planck \cite{PlanckVI_2018})\\
    \end{tabular}
    \caption{Summary of limits on neutrino magnetic moments.  Adapted from Table III of Ref.\ \cite{2015RvMP...87..531G}.
    }
    \label{tab:limits}
\end{table}

We note two points about our work which is unique to Dirac neutrinos but not anomalous magnetic moments in the context of cosmology.  The first is the fact that there must exist 3 eigenstates for the inactive states with mass eigenvalues identical to the active neutrinos.  When considering the energy density of a new low-mass particle in the Early Universe, Dirac neutrinos come with a factor of 3 attached and as a result increase the energy density over that of a single neutral fermion [see Fig.\ (21) in Ref.\ \cite{cmbs4_science_book}; and Ref.\ \cite{2021PhLB..82336736A}].  Second and related to the first, the inactive neutrinos have non-zero masses -- negligible at early times but not at late.  Structure growth and neutrino free-streaming are indeed dependent on the population of the inactive neutrinos, although those states cannot be thermally populated.  A corollary of this result is that the inactive states must have different temperatures or spectra than the actives.
Together, the low-mass and extra states of Dirac neutrinos give two methods to probe the neutrino spectra and search for new physics in cosmology.

We have examined the implications of Dirac neutrinos with anomalous magnetic moments on early and late-time cosmology in this work; and is a follow-up to the Majorana case of Ref.\ \cite{Vassh_tdecoup}.
In a standard seesaw mechanism  \cite{PhysRevLett.44.912,Balantekin:2018azf}, the three active states are Majorana neutrinos; the three sterile states have mass eigenvalues much heavier than the active states and cannot be probed by current cosmological observations.  However, there exists a possibility that those three sterile states could have mass eigenvalues nearly degenerate with the active ones.  This possibility of neutrinos being ``pseudo-Dirac'' has been studied in the case of the diffuse supernova background \cite{2020PhRvD.102l3012D} and mentioned in the case of early-time cosmology \cite{2009PhRvD..80g3007D}.  If this mass model holds and neutrinos have anomalous magnetic moments, then they are Majorana particles and the analysis of BBN in Ref.\ \cite{Vassh_tdecoup} would apply. 
In addition, the analyses on early and late time cosmological energy density in this work would also be relevant.
Although such a hybrid situation is intriguing, Ref.\ \cite{Vassh_tdecoup} showed that the magnetic moment needs to be $\sim10^{-10}\mu_B$ to influence the neutron-to-proton rates (and subsequent abundances) through altered neutrino spectra.  If additional non-active or sterile states can be populated via an anomalous magnetic moment, then Fig.\ \ref{fig:dneff_vs_mu} shows that the \neff would be nearly 6.0 and ruled out by current cosmological parameter estimation.
For early-time cosmology, pseudo-Dirac cannot be distinguished from uniquely Dirac via anomalous magnetic moments.

Where there might be a difference between these two mass models is interpreting \summnu from large-scale-structure growth.
If $\kappa<10^{-10}\mu_B$ and the mass eigenvalues for the sterile states are nearly degenerate, then this situation is a close reproduction to the situation studied in Sec.\ \ref{sec:later}. To be precise, we would need to slightly alter the summations over the sterile states in Eq.\ \eqref{eq:vth_av1} to account for different masses, although this should not make a significant difference if the sterile mass eigenvalues are nearly degenerate with the active ones.
If, however, the sterile masses are \emph{smaller} than the active ones, there would be a contribution to \neff but not to the dark-matter contribution at late times (see Fig.\ \ref{fig:kfs_vs_m1}), thereby changing the free-streaming scale for the active neutrinos.  In addition, lighter sterile states also introduce the possibility of active neutrino decays which alter the dynamics in late-time cosmology \cite{Beacom:2004yd, Serpico:2007pt, Farzan:2015pca}.
For this scenario of light sterile states and anomalous magnetic moments, pseudo-Dirac and uniquely Dirac give different predictions in late-time cosmology.

Finally, we comment on an often quoted result in the literature, colloquially referred to as the ``Kayser confusion theorem'' \cite{1982PhRvD..26.1662K,Kayser:1981nw}.
The theorem points out that other neutrino properties could lead to similar effects as the magnetic moment in scattering processes.  Specifically, a Dirac magnetic moment could be confused with a Majorana anapole moment [see the $f_M$ and $f_A$ terms in Eq.\ \eqref{eq:form_factor}].  
In this connection it is worthwhile to point out that 
cosmology differs from particle-beam experiments in one key aspect.  In brief, the cosmological parameters \neff and \summnu give a measure of the energy density, i.e., they indicate which states are populated by particles and have distinct manifestations for Majorana versus Dirac character. 
Conversely, particle-beam experiments measure cross sections and have difficulties discerning between the two characters as discussed in Refs.\ \cite{1982PhRvD..26.1662K,Kayser:1981nw}.
We do not advocate for using one method over the other.  Rather, both should be pursued as they complement one another in probing the nature of neutrino interactions.

\section*{Acknowledgements} 

The authors thank Volker Koch, Volodymyr Vovchenko, Nicole Vassh, George Fuller, James Kneller, Gail McLaughlin, and Amol Patwardhan for useful discussions.
EG is supported in part by the Department of Energy Office of Nuclear Physics award DE-FG02-02ER41216, and by the National Science Foundation grant No.\ PHY-1430152 (Joint Institute for Nuclear Astrophysics Center for the Evolution of the Elements). 
ABB is supported in part by the National Science Foundation Grant  PHY-2108339 at University of Wisconsin Madison.
ABB and EG acknowledge supported in part by the National Science Foundation Grants No. PHY-1630782 and PHY-2020275 (Network for Neutrinos Astrophysics and Symmetries).
This research used resources provided by the Los Alamos National Laboratory Institutional Computing Program, which is supported by the U.S. Department of Energy National Nuclear Security Administration under Contract No. 89233218CNA000001.

\appendix

\section{Differential Cross Sections with Magnetic Moment Vertex}
\label{app:a}

Here we give differential cross sections for the various scattering and annihilation processes with fermions and bosons.  The differential cross sections are functions of Mandelstam variable $t=(p_1-p_3)^2$ for the reaction $1+2\leftrightarrow 3+4$, and depend on Mandelstam variable $s=(p_1+p_2)^2$, the neutrino magnetic moment strength $\mu$, the effective in-medium photon mass $m_\gamma$ from Eq.\ \eqref{debye}, and the vacuum mass of the charged boson/fermion.  We calculate the integrated cross sections using
\begin{equation}
  \sigma(s) = \int_{t_{\rm min}}^{t_{\rm max}}dt\, \frac{d\sigma}{dt},
\end{equation}
where the limits of integration are
\begin{align}
  t_{\rm min} &= -\frac{(s-m_i^2)^2}{s} \\
  t_{\rm max} &= 0
\end{align}

\subsection{Fermions}

\begin{figure}
  \includegraphics[width=0.5\columnwidth]{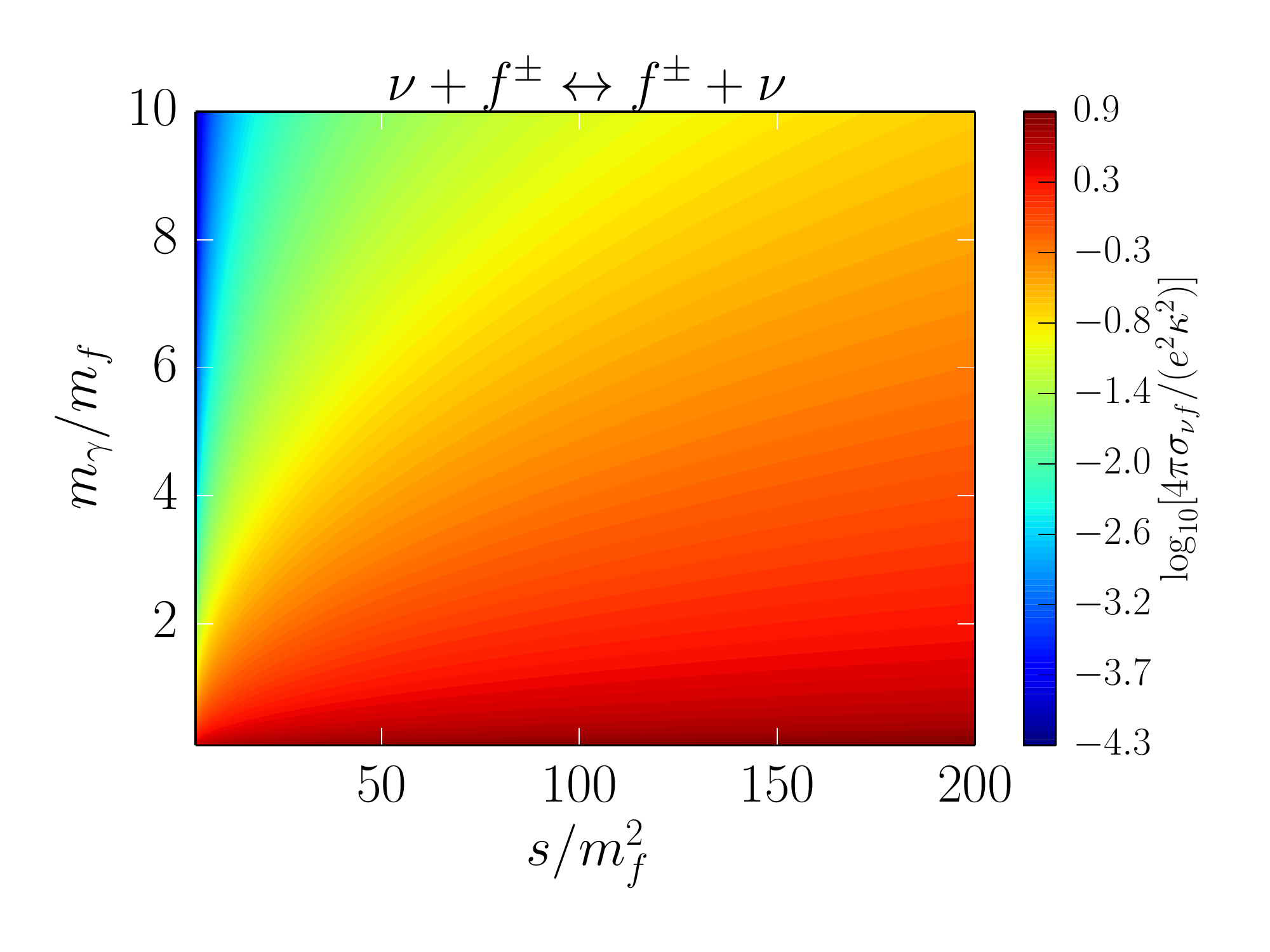}
  \caption{\label{fig:mg_s_cont_sigma_f} $m_\gamma$ versus kinetic variable $s$
  plotted at contours of constant $\sigma_{\nu f}$ for $\nu+f^\pm\leftrightarrow
  f^\pm+\nu$ elastic scattering channel [see Eq.\ \eqref{app:nmmxc}].  We take $q_f=1$ in this figure.
  }
\end{figure}

Equation \eqref{nmmxc} gave the differential cross section for the scattering of neutrinos from charged fermions of mass $m_f$
and charge-coefficient $q_f$
\begin{equation}
\label{app:nmmxc}
  \left(\frac{d \sigma}{dt}\right)_{\nu f}
  = \frac{\pi q_f^2 \alpha^2}{m_e^2}
  \mu^2 \frac{t}{(t-m_{\gamma}^2)^2} \frac{s+t-m_f^2}{s-m_f^2}.
\end{equation}
Figure \ref{fig:mg_s_cont_sigma_f} shows a contour plot of the total cross section in the $m_\gamma$ versus $s$ plane.
The magnetic moment contribution to the neutrino-antineutrino annihilation differential cross section into charged fermion-antifermion pairs each with mass $m_f$ is given by 
\begin{equation}
  \left(\frac{d \sigma}{dt}\right)_{f\overline{f}} = \left(\frac{ 2\pi q_f^2
  \alpha^2}{m_e^2} \right) \mu^2 \frac{1}{s(s-m_{\gamma}^2)^2}  (t+s -m_f^2)(m_f^2-t).
\end{equation}

\subsection{Bosons}

\begin{figure}
  \includegraphics[width=0.5\columnwidth]{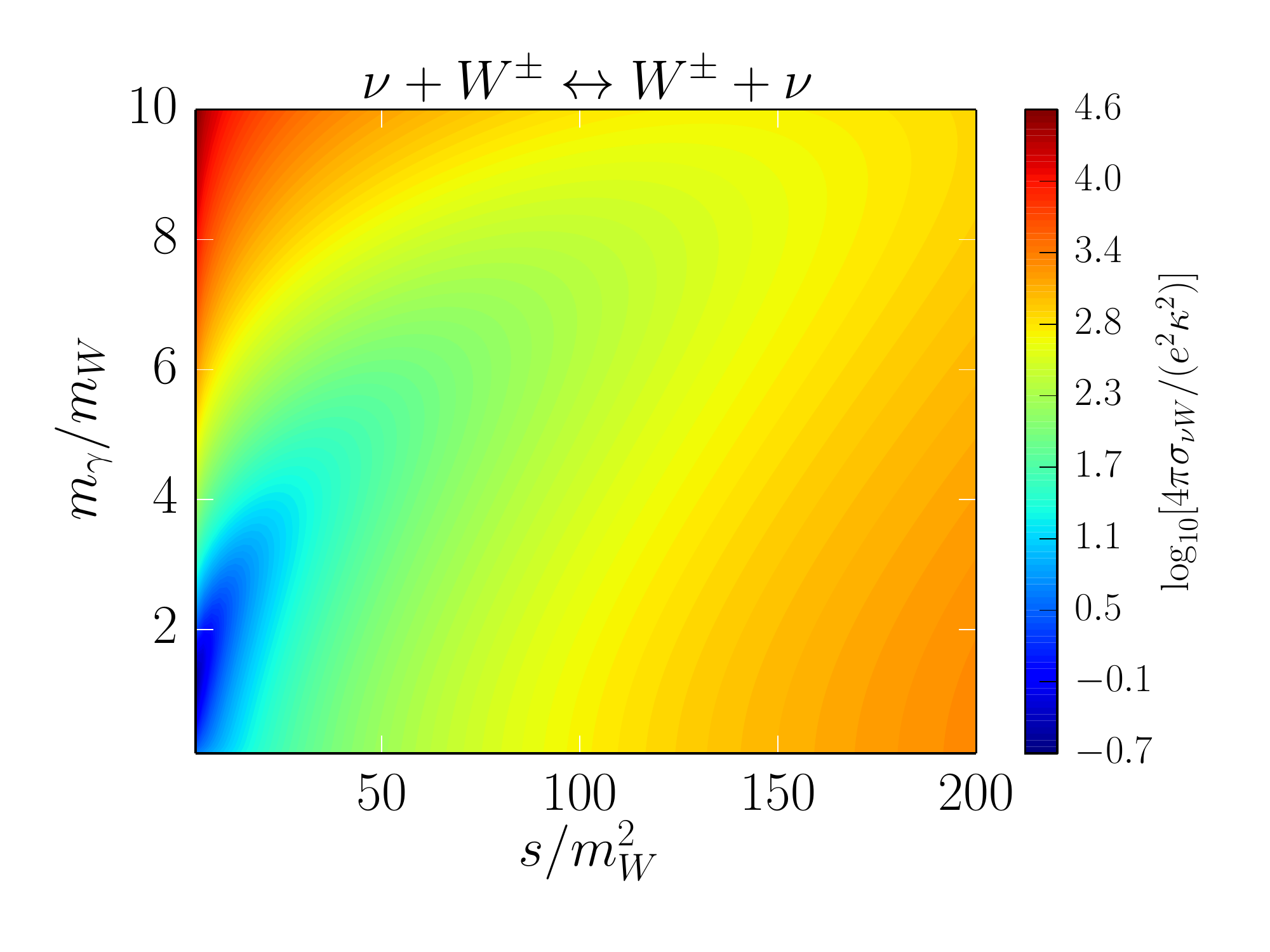}
  \caption{\label{fig:mg_s_cont_sigma_b} $m_\gamma$ versus kinetic variable $s$
  plotted at contours of constant $\sigma_{\nu W}$ for $\nu+W^\pm\leftrightarrow
  W^\pm+\nu$ elastic scattering channel [see Eq.\ \eqref{eq:nuW}].
  }
\end{figure}

The only boson particle we consider is $W^{\pm}$.  The cross section for the magnetic moment contribution to the scattering of neutrinos from $W$-bosons with mass $m_W\simeq80.4\,{\rm GeV}$ is given by 
\begin{equation}\label{eq:nuW}
\left(\frac{d \sigma}{dt}\right)_{\nu W} = \frac{\pi \alpha^2}{m_e^2}  \mu^2 
\frac{t}{(t-m_{\gamma}^2)^2} \left[ \left( 1 - \frac{t}{3m_W^2} + \frac{ t^2}{4m_W^4} \right)  \frac{s+t- m_W^2}{s-m_W^2}  +  \left(-\frac{5}{12} - \frac{t}{16m_W^2} + \frac{ t^2}{16m_W^4} \right) \frac{t^2}{(s-m_W^2)^2} \right] .
\end{equation}
Figure \ref{fig:mg_s_cont_sigma_b} gives contours of total cross section in the $m_\gamma$ versus $s$ plane.
Finally, the magnetic moment contribution to the neutrino-antineutrino annihilation cross section into $W^+$-$W^-$ pairs is given by 
\begin{equation}
\left(\frac{d \sigma}{dt}\right)_{W^+W^-} = \frac{e^2 \kappa^2}{16 \pi} \frac{1}{s(s-m_{\gamma}^2)^2} \left[ (s+2t -m_W^2)^2 \left( 3 - \frac{s}{m_W^2} + \frac{s^2}{4 m_W^4} \right) + 8s \left(-s + \frac{s^2}{4m_W^2} \right)    \right] .
\end{equation}
In calculating the cross sections involving $W$-bosons we ignored the four-boson couplings since the associated amplitudes are suppressed by another order of the magnetic moment. 

\subsection{Hadrons}

\begin{figure}
  \includegraphics[width=0.5\columnwidth]{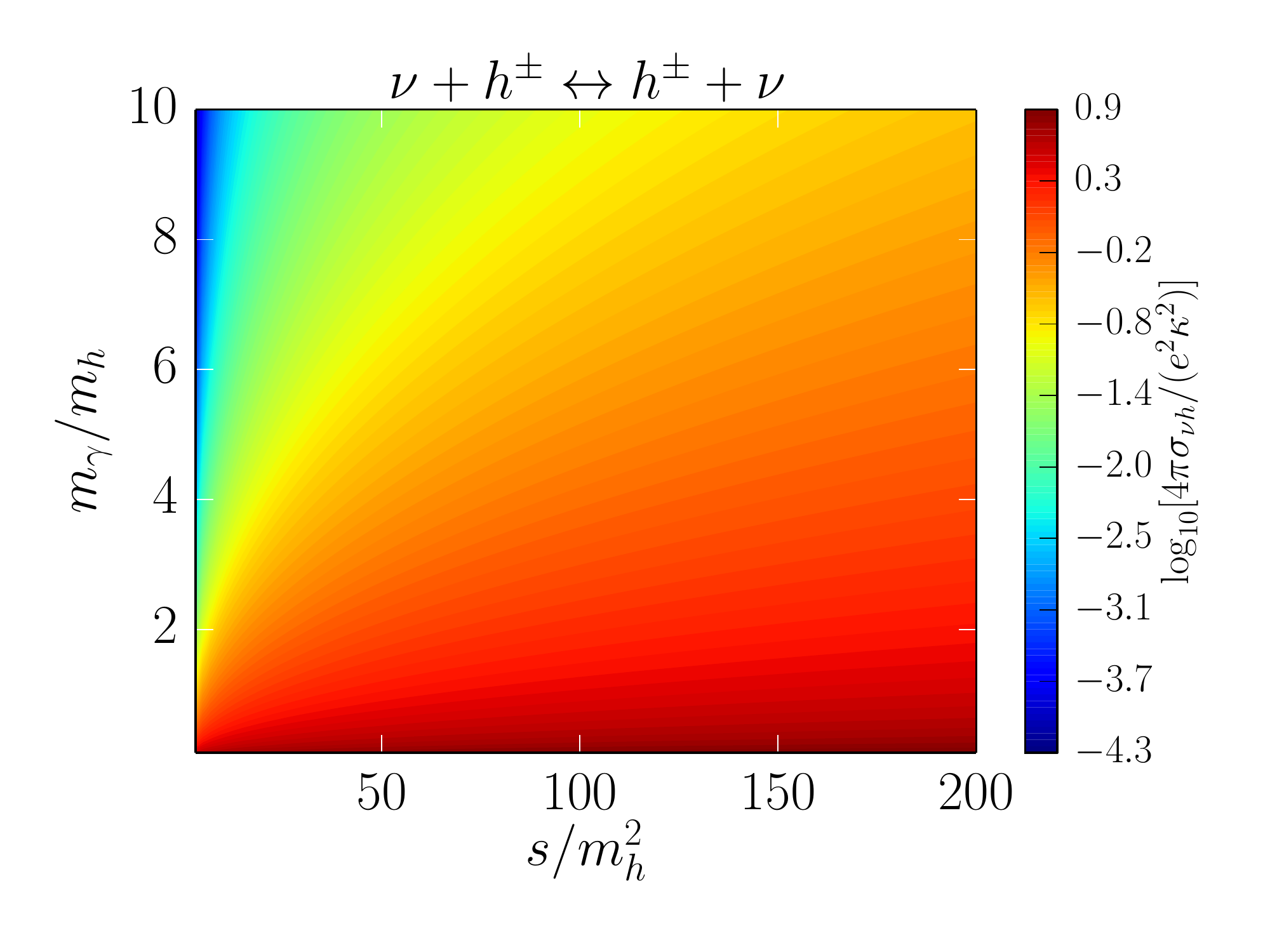}
  \caption{\label{fig:mg_s_cont_sigma_h} $m_\gamma$ versus kinetic variable $s$
  plotted at contours of constant $\sigma_{\nu h}$ for $\nu+h^\pm\leftrightarrow
  h^\pm+\nu$ elastic scattering channel [see Eq.\ \eqref{eq:nuh}].
  }
\end{figure}

The cross section for scattering on scalar charged hadrons with mass $m_h$ is
\begin{equation}\label{eq:nuh}
  \left(\frac{d\sigma}{dt}\right)_{\nu h} = \frac{e^2\kappa^2}{4\pi}\frac{t}{(t-m_\gamma^2)^2}\left[
  1+\frac{t}{s-m_h^2}+\frac{t^2}{4(s-m_h^2)^2}\right].
\end{equation}
Figure \ref{fig:mg_s_cont_sigma_h} gives contours of total cross section in the $m_\gamma$ versus $m_h$ plane.
We take the annihilation cross section into scalar hadron-antihardron pairs to be zero.
In the case of charged vector hadrons, we ignore the contributions to the scattering rates, and so do not provide the scattering and annihilation differential cross sections.  Although these rates would be non-zero, we estimate only a small error as the vector hadrons have large masses and do not appear in appreciable numbers at temperatures below the QHT (see rows 7 and 9 in Table \ref{tab:hadrons}).

\section{Thermally-averaged Cross Sections}
\label{app:b}

\subsection{Elastic Scattering}

We use the thermally-averaged product of $\sigma$ and $v_{\rm Mol}$, denoted $\langle\sigma v_{\rm Mol}\rangle$, to calculate scattering rates between neutrinos and other particles via the magnetic-moment vertex.  The formula for the thermal average is the following
\begin{align}
  \langle\sigma v_{\rm Mol}\rangle &= \dfrac{\frac{g_1g_2}{(2\pi)^6}\int d^3p_1\frac{1}{e^{E_1/T}+1}\int d^3p_2\,\sigma v_{\rm Mol}\frac{1}{e^{E_2/T}\pm1}}
  {\frac{g_1g_2}{(2\pi)^6}\int d^3p_1\frac{1}{e^{E_1/T}+1}\int d^3p_2\frac{1}{e^{E_2/T}\pm1}}\\
  &= \frac{g_1g_2}{(2\pi)^6n_1n_2}\int d^3p_1\frac{1}{e^{E_1/T}+1}\int d^3p_2\,\sigma v_{\rm Mol}\frac{1}{e^{E_2/T}\pm1},\label{eq:sv1}
\end{align}
where we have assumed equilibrium distributions and ignored the Pauli blocking/Bose enhancement of the products.
Particle 1 is the neutrino with zero rest mass, and particle 2 is the
scattering target with rest mass $m$.  The $\pm1$ in the distribution function
for the 2nd particle corresponds to either fermions $(+)$ or bosons $(-)$.  Both $\sigma$ and $v_{\rm Mol}$ are given in terms of Mandelstam variable
$s$, particle 2 mass $m$, and the in-medium photon mass $m_\gamma$.

With a change in variables and using $v_{\rm Mol}=(s-m^2)/2E_1E_2$ \cite{1991NuPhB.360..145G}, we can reduce the expression in Eq.\ \eqref{eq:sv1} to a double integral.  For fermions, that expression is
\beq
  \langle\sigma v_{\rm Mol}\rangle_{\rm FD}
  = \frac{2g_1g_2\pi^2T^6}{(2\pi)^6n_1n_2}\int\limits_{\epsilon_m^2}^{\infty} d\epsilon_s\,(\epsilon_s-\epsilon_m^2)\,\sigma
  \int\limits_{\sqrt{\epsilon_s}}^{\infty} d\epsilon_+\,
  \frac{1}{e^{\epsilon_+}-1}
  \left\{\beta + \ln\left[\dfrac{1+2e^{(-\beta-\epsilon_+)/2}\cosh\left(\frac{\alpha}{2}\right)+e^{-\beta-\epsilon_+}}
  {1+2e^{(\beta-\epsilon_+)/2}\cosh\left(\frac{\alpha}{2}\right)+e^{\beta-\epsilon_+}}\right]\right\}, \label{eq:sv6}
\eeq
where the $\epsilon$ notation denotes an energy quantity normalized by the appropriate power of $T$, namely $\epsilon_s=s/T^2$, $\epsilon_+=E_+/T$, $\epsilon_m=m/T$, and $\epsilon_\gamma=m_\gamma/T$.  We rewrite $\sigma$ as a function of $\epsilon_s$, $\epsilon_m$, $\epsilon_\gamma$ and $T$.  In Eq.\ \eqref{eq:sv6}, we have also defined new quantities for ease in writing
\beq
  \alpha = \epsilon_+\frac{\epsilon_m^2}{\epsilon_s},\quad
  \beta = \frac{\epsilon_s-\epsilon_m^2}{\epsilon_s}\sqrt{\epsilon_+^2 - \epsilon_s}.
\eeq
Our expression in Eq.\ \eqref{eq:sv6} is the same as Eq.\ (B6) in Ref.\ \cite{Vassh_tdecoup} where we have corrected a few typographical errors.

For bosons, the thermal average for elastic scattering is 
\beq
  \langle\sigma v_{\rm Mol}\rangle_{\rm BE}
  = \frac{2g_1g_2\pi^2T^6}{(2\pi)^6n_1n_2}\int\limits_{\epsilon_m^2}^{\infty} d\epsilon_s\,(\epsilon_s-\epsilon_m^2)\,\sigma
  \int\limits_{\sqrt{\epsilon_s}}^{\infty} d\epsilon_+\,
  \frac{1}{e^{\epsilon_+}+1}
  \ln\left[\dfrac{\sinh\left(\frac{\alpha}{2}\right) + \sinh\left(\frac{\epsilon_+ + \beta}{2}\right)}
  {\sinh\left(\frac{\alpha}{2}\right) + \sinh\left(\frac{\epsilon_+-\beta}{2}\right)}\right],
\eeq
with the same notation as Eq.\ \eqref{eq:sv6}.

\subsection{Annihilation Scattering}

For the annihilation channels, the expression for $\langle\sigma v_{\rm Mol}\rangle$ is the same for either boson or fermion pairs with mass $m$, as we average over the initial neutrino-antineutrino distributions.  The result is the same expression as Eq.\ \eqref{eq:sv6} except with a different threshold value of $s$ and massless reactants
\beq
  \langle\sigma v_{\rm Mol}\rangle_{\rm ann}
  = \frac{4g_1g_2\pi^2T^6}{(2\pi)^6n_1n_2}\int\limits_{4\epsilon_m^2}^{\infty} d\epsilon_s\,\epsilon_s\,\sigma
  \int\limits_{\sqrt{\epsilon_s}}^{\infty} d\epsilon_+\,
  \frac{1}{e^{\epsilon_+}-1}
  \ln\left[\dfrac{\cosh\left(\frac{\epsilon_+ + \beta}{4}\right)}
  {\cosh\left(\frac{\epsilon_+-\beta}{4}\right)}\right],
\eeq
where $\beta = \sqrt{\epsilon_+^2 - \epsilon_s}$.

\section{Treatment of Quark-Hadron Transition in the Early Universe}
\label{app:qht}

We have considered a range of models of anomalous magnetic moments which include decoupling of the inactive Dirac states in the $T\sim100\,{\rm MeV}$ range.  Decoupling in this range is complicated by the transition from free quarks and gluons to bound hadrons in an expanding and cooling universe.  As a result, we model this epoch using a smooth crossover from a quark-gluon equation of state to one dominated by hadrons.

At high temperatures, we approximate the quark-gluon ($qg$) component as an ideal gas with negligible chemical potential.
The $qg$ component includes the six quark and gluon degrees of freedom with appropriate degeneracy factors.
Conversely, at low temperature, we also approximate the hadron ($h$) component as an ideal gas with negligible chemical potential.
We use the lightest hadrons shown in Table \ref{tab:hadrons}.  The next heaviest hadrons after the $K^{\ast 0}(896)$ states are protons and neutrons.  We have verified that excluding those baryons from the hadron component do not alter any of our results.

\begin{table}
    \centering
    \begin{tabular}{ccccc}
         Name & mass (MeV) & charge & degeneracy \\
        \hline
        $\pi^0$ & 135.0 & 0 & 1\\
        $\pi^+$ & 140.0 & 1 & 1\\
        $K^+$ & 494.0 & 1 & 1\\
        $K^0$ & 498.0 & 0 & 2\\
        $\eta^0$ & 548.0 & 0 & 1\\
        $\rho^0$ & 775.0 & 0 & 3\\
        $\rho^+$ & 775.0 & 1 & 3\\
        $\omega^0$ & 783.0 & 0 & 3\\
        $K^{\ast +}(892)$ & 892.0 & 1 & 3\\
        $K^{\ast 0}(896)$ & 896.0 & 0 & 6
    \end{tabular}
    \caption{Table of hadrons used in the early universe for this work.  First column is name/symbol of the particle.  Second, third, and fourth columns are mass (MeV), charge, and degeneracy (respectively).  All positively charged hadrons have negatively charged partners.  All particles are bosons.
    }
    \label{tab:hadrons}
\end{table}

We use the combination of $qg$ and $h$ components to calculate thermodynamic quantities and the Debye screening length.
We denote the pressure, entropic density, number density, and energy density for the $qg$ component as $P_{qg}$, $s_{qg}$, $n_{gq}$ and $\rho_{qg}$, respectively.  For the hadrons, we replace the $qg$ subscript with an $h$.
To weight the contributions from the two seas when both components are present, we use a switching function following a prescription from Ref.\ \cite{2014PhRvC..90b4915A}
\begin{align}
  S(T,\mu) &= \exp[-\theta(T,\mu)]\\
  \theta(T,\mu) &= \left[\left(\frac{T}{T_0}\right)^r + \left(\frac{\mu}{\mu_0}\right)^r\right]^{-1}.
\end{align}
The switching function uses data calculated with lattice QCD \cite{Borsanyi:2010cj} to fit the parameters $T_0, \mu_0$, and $r$.
We use $r=4$, $T_0=145.33\,{\rm MeV}$, and $\mu_0=3\pi T_0$ from the first row of Table I in Ref.\ \cite{2014PhRvC..90b4915A}.  $\mu$ is the chemical potential.

\begin{figure}
  \includegraphics[width=0.5\columnwidth]{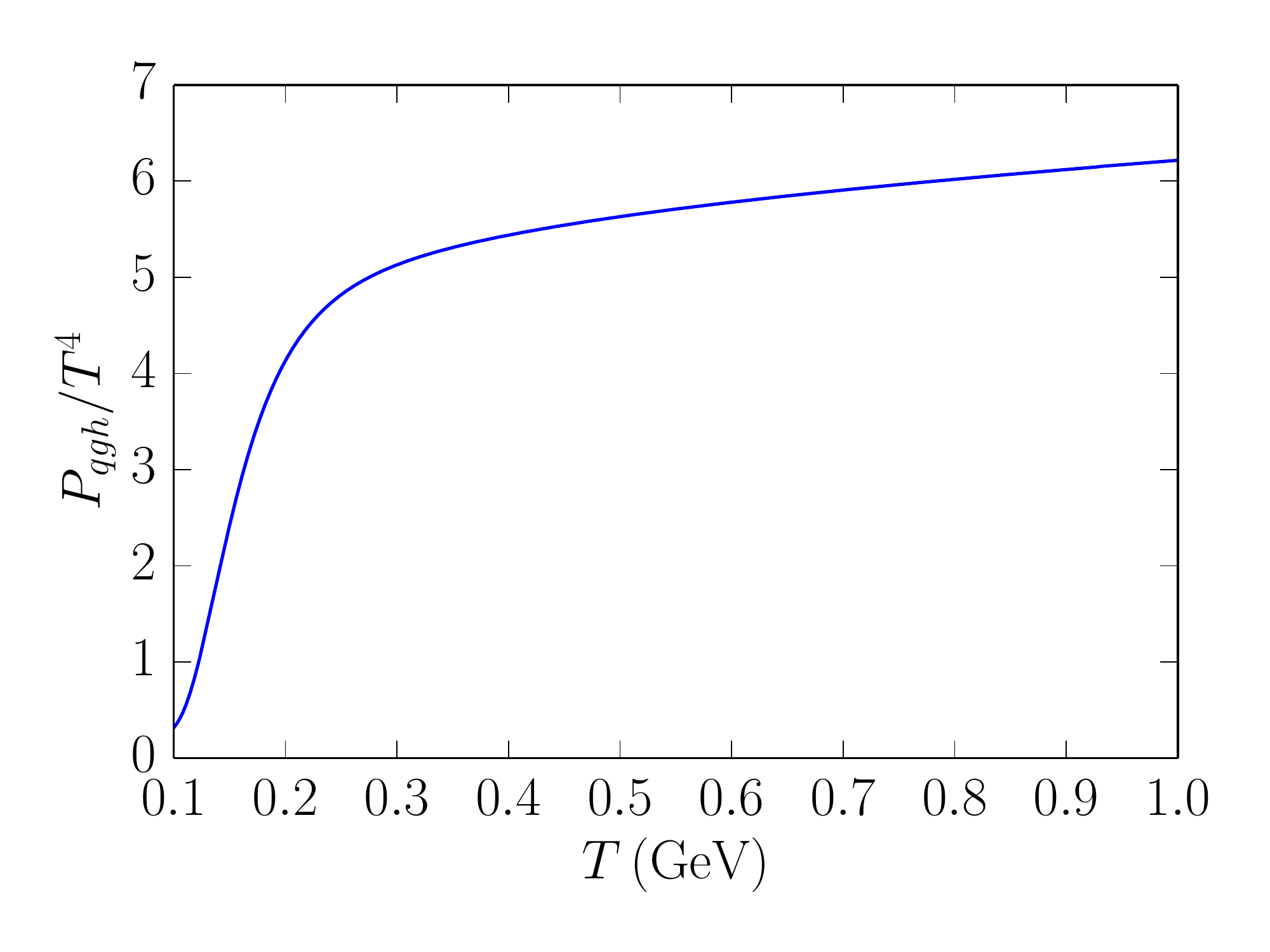}
  \caption{\label{fig:p_v_t_qhpt} Pressure of quark-hadron components versus
  temperature [see Eq.\ \eqref{eq:p_qgh}].  At high temperature, the system approaches that of an ideal gas
  of quarks and gluons.  At low temperature, the system approaches that of an
  ideal gas of hadrons.  For temperatures in the 100 MeV range, results are used
  from lattice QCD.
  }
\end{figure}

According to the procedure in Ref.\ \cite{2014PhRvC..90b4915A}, we apply the switching function directly to the pressure
\begin{equation}\label{eq:p_qgh}
  P_{qgh} = S(T,\mu)P_{qg} + [1-S(T,\mu)]P_{h},
\end{equation}
where $P_{qgh}$ is the total pressure supplied by the quarks, gluons, and hadrons.  Figure \ref{fig:p_v_t_qhpt} shows $P_{qgh}/T^4$ as a function of $T$ for $\mu=0$.  The increase in $P_{qgh}$ between $T=100\,{\rm MeV}$ and $T=200\,{\rm MeV}$ is due to the appearance of the $qg$ degrees of freedom and concomitant disappearance of the $h$ degrees of freedom. The expressions for $s$, $n$, and $\rho$ follow from derivatives of the pressure
\begin{align}
  s_{qgh} &= Ss_{qg} + (1-S)s_h +
  S\frac{r\theta^2}{T}\left(\frac{T}{T_0}
  \right)^r(P_{qg} - P_h),\\
  n_{qgh} &= Sn_{qg} + (1-S)n_h +
  S\frac{r\theta^2}{\mu}\left(\frac{\mu}{\mu_0}
  \right)^r(P_{qg} - P_h),\label{eq:switch_n}\\
  \rho_{qgh} &= Ts_{qgh} - P_{qgh} + \mu n_{qgh}.
\end{align}
When calculating the inverse square of the Debye length in Eq.\ \eqref{debye2}, the relevant quantity is the derivative of the number density with respect to $\mu$.
We would need to take the derivative of Eq.\ \eqref{eq:switch_n} with respect to $\mu$ to calculate $m_\gamma^2$ during the QHT with the switching function.  However, in the $CP$-symmetric conditions of the early universe, all derivatives of the switching function with respect to $\mu$ are zero for $\mu=0$.  Hence, the expression for the contribution to $m_\gamma^2$ from the quark-gluon-hadron components are
\begin{equation}
  m_{\gamma,\,qgh}^2 = 4\pi\alpha\left\{S(T,\mu=0)\sum_j q_j^2 \frac{\partial}{\partial \mu} [ n_j^{(-)} - n_j^{(+)}]
  + [1-S(T,\mu=0)]\sum_k q_k^2 \frac{\partial}{\partial \mu} [ n_k^{(-)} - n_k^{(+)}]\right\},
\end{equation} 
where the first summation is over quark pairs and the second summation is over charged hadron pairs.

\bibliography{MM_dirac_bib}

\end{document}